\documentclass{article}

\usepackage{arxiv}

\usepackage[utf8]{inputenc} % allow utf-8 input
\usepackage[T1]{fontenc}    % use 8-bit T1 fonts
\usepackage{hyperref}       % hyperlinks
\usepackage{url}            % simple URL typesetting
\usepackage{booktabs}       % professional-quality tables
\usepackage{amsfonts}       % blackboard math symbols
\usepackage{nicefrac}       % compact symbols for 1/2, etc.
\usepackage{microtype}      % microtypography
\usepackage{lipsum}
\usepackage{graphicx}
\graphicspath{ {./images/} }

\usepackage{multirow}
\usepackage{algorithm}
\usepackage{algpseudocode}
\usepackage{amsmath}
\usepackage[sort&compress]{natbib}
\usepackage{rotating}

\title{Deep Synthetic Cross-Project Approaches for Software Reliability Growth Modeling}

\author{
 Taehyoun Kim \\
  School of Computing\\
  Korea Advanced Institute of Science and Technology (KAIST)\\
  Daejeon, Republic of Korea\\
  \texttt{tae\_hyoun@kaist.ac.kr} \\
   \And
 Duksan Ryu \\
  Department of Software Engineering\\
  Jeonbuk National University\\
  Jeonju-si, Jeonbuk-do, Republic of Korea\\
  \texttt{duksan.ryu@jbnu.ac.kr} \\
  \And
 Jongmoon Baik \\
  School of Computing\\
  Korea Advanced Institute of Science and Technology (KAIST)\\
  Daejeon, Republic of Korea\\
  \texttt{jbaik@kaist.ac.kr} \\
}

\begin{document}
\maketitle
\begin{abstract}
Software Reliability Growth Models (SRGMs) are widely used to predict software reliability based on defect discovery data collected during testing or operational phases. However, their predictive accuracy often degrades in data-scarce environments, such as early-stage testing or safety-critical systems. Although cross-project transfer learning has been explored to mitigate this issue by leveraging data from past projects, its applicability remains limited due to the scarcity and confidentiality of real-world datasets.
To overcome these limitations, we propose Deep Synthetic Cross-project SRGM (DSC-SRGM), a novel approach that integrates synthetic data generation with cross-project transfer learning. Synthetic datasets are generated using traditional SRGMs to preserve the statistical characteristics of real-world defect discovery trends. A cross-correlation-based clustering method is applied to identify synthetic datasets with patterns similar to the target project. These datasets are then used to train a deep learning model for reliability prediction.
The proposed method is evaluated on 60 real-world datasets, and its performance is compared with both traditional SRGMs and cross-project deep learning models trained on real-world datasets. DSC-SRGM achieves up to 23.3\% improvement in predictive accuracy over traditional SRGMs and 32.2\% over cross-project deep learning models trained on real-world datasets. However, excessive use of synthetic data or a naive combination of synthetic and real-world data may degrade prediction performance, highlighting the importance of maintaining an appropriate data balance. These findings indicate that DSC-SRGM is a promising approach for software reliability prediction in data-scarce environments.
\end{abstract}

% keywords can be removed
\keywords{software reliability growth model \and synthetic data generation \and cross-project transfer learning \and software reliability prediction}

\section{Introduction}
Software Reliability Growth Models (SRGMs) have been widely adopted to analyze and predict software reliability based on defect discovery data collected during testing or operational phases~\cite{wood1996software}. By estimating the cumulative number of defects over time, SRGMs help software engineers evaluate release readiness and allocate resources efficiently~\cite{okumoto1979optimum, rani2021entropy}.

Traditional SRGMs are mathematical models based on probabilistic assumptions regarding defect occurrence and detection~\cite{shooman1972probabilistic, goel1985software}. Representative models, including the Goel-Okumoto (GO) model~\cite{goel1979software, goel1979time}, the Yamada Delayed S-Shaped (YDSS) model~\cite{yamada2009s}, the Inflection S-Shaped (ISS) model~\cite{ohba1984inflection}, and the Generalized Goel (GG) model~\cite{goel1982software}, have been extensively utilized in both academic research and industrial applications. While these models provide meaningful insights, their predictive accuracy depends on how well the observed defect discovery data conform to their statistical assumptions~\cite{kapur2011software}. Therefore, selecting an appropriate model for a given dataset requires careful evaluation of model assumptions and data characteristics~\cite{sharma2010selection}. Furthermore, these models require sufficient data to ensure stable parameter estimation~\cite{stringfellow2002empirical, jabeen2019improved}. When the available data are sparse or noisy, parameter estimates become unstable, leading to significant degradation in predictive accuracy~\cite{li1993enhancing, cinque2017debugging}.

To address these limitations, machine learning (ML) and deep learning (DL) approaches have been explored for software reliability modeling~\cite{yang2010generic, pai2006software, das2011failure, karunanithi1992prediction, cai2001neural, wang2018software, behera2018software, gusmanov2019cnn, yangzhen2017software, wu2020hybrid, yang2024software, kumar2012empirical, rani2018neural}. Unlike traditional SRGMs, these approaches do not rely on explicit probabilistic assumptions and can learn complex, non-linear patterns directly from defect discovery data. Various models have been investigated, including Support Vector Machines (SVM)~\cite{yang2010generic, pai2006software, das2011failure}, Artificial Neural Networks (ANN)~\cite{karunanithi1992prediction, cai2001neural}, Recurrent Neural Networks (RNN)~\cite{wang2018software, behera2018software}, Long Short-Term Memory (LSTM)~\cite{gusmanov2019cnn, yangzhen2017software}, and Gated Recurrent Units (GRU)~\cite{wu2020hybrid, yang2024software}. These models have shown promising results in capturing temporal dependencies in defect discovery data. However, similar to traditional SRGMs, their effectiveness is limited in data-scarce environments. ML- and DL-based approaches require substantial training data to produce generalizable predictions. When data are limited, these models are prone to overfitting, which reduces their ability to generalize to unseen cases.

A common limitation of both traditional and data-driven models is their dependence on sufficient defect discovery data for accurate prediction. However, data scarcity frequently arises from several factors encountered in software projects. During early testing stages, defect discovery is inherently limited, resulting in incomplete datasets~\cite{xie1997practical}. In safety-critical or high-assurance systems, defects are rare by design, which makes data collection even more difficult~\cite{azturk2021ensemble, zhang2024bootstrap}. These constraints significantly impact the predictive accuracy of both traditional and data-driven models.

Cross-project transfer learning has been explored as a potential solution to mitigate the effects of target data scarcity~\cite{san2021deep, nagaraju2022adaptive}. This approach leverages defect discovery data from past projects to improve the reliability prediction of a target project. While this approach has shown meaningful results, its effectiveness is also constrained by the scarcity and confidentiality of publicly available datasets. Most organizations treat defect discovery data as proprietary and do not share it publicly~\cite{wood1996predicting, karunanithi1992prediction, karunanithi1991prediction}. Even when such datasets are accessible~\cite{musa1979software, lyu1996handbook, mivcko2022applicability}, they are often not sufficiently diverse or extensive to support robust cross-project learning. These datasets tend to be small, outdated, or biased toward specific software types, which limits their applicability to broader software projects~\cite{kim2024enhancing}.

To overcome these limitations, this study aims to enable accurate software reliability prediction even when both target and past project data are limited. We propose Deep Synthetic Cross-project SRGM (DSC-SRGM), a novel approach that integrates synthetic data generation with cross-project transfer learning. Synthetic datasets substitute for real-world past project data in the training process. These synthetic datasets are generated using well-established traditional SRGMs to preserve the statistical characteristics of real-world defect discovery trends. A similarity score-based clustering method using cross-correlation is then applied to identify synthetic datasets with patterns similar to the target project. These selected datasets are used to train a deep learning model for reliability prediction. By integrating synthetic data generation with cross-project transfer learning, DSC-SRGM enables reliable and generalizable prediction in data-scarce environments. To the best of our knowledge, this is the first approach to leverage traditional SRGMs for generating synthetic data in the context of cross-project software reliability modeling.

We evaluated the effectiveness of DSC-SRGM through extensive experiments on 60 real-world software projects. The results show that DSC-SRGM outperforms both traditional SRGMs and cross-project deep learning models trained on real-world datasets. We also investigated how combining synthetic and real-world data, or varying the amount of synthetic data, affects predictive performance. The findings reveal that combining the two data types does not necessarily improve accuracy, and that excessive use of synthetic data may degrade performance due to overfitting or distortion of defect discovery patterns. These results highlight the importance of not only generating synthetic data but also selecting and using it in a balanced and informed manner.

The main contributions of this study are summarized as follows:
\begin{itemize}
  \item DSC-SRGM is introduced as a novel software reliability modeling approach that integrates synthetic data generation with cross-project transfer learning to address the challenge of data scarcity.

  \item The effectiveness of DSC-SRGM is validated through extensive experiments on 60 real-world software projects, with results showing consistent performance improvements over both traditional SRGMs and cross-project deep learning models trained on real-world datasets.

  \item Additional analyses demonstrate that combining real-world and synthetic data does not necessarily improve performance, and that excessive synthetic data may degrade predictive accuracy, emphasizing the importance of balanced data usage.
\end{itemize}

The remainder of this paper is organized as follows. Section~\ref{sec:related} reviews related work on cross-project software reliability modeling and summarizes publicly available defect discovery datasets. Section~\ref{sec:method} presents the proposed DSC-SRGM methodology, including synthetic data generation, similarity score-based clustering, and deep learning-based prediction. Section~\ref{sec:setup} describes the experimental setup, datasets, and evaluation metrics. Section~\ref{sec:results} presents the experimental results, followed by ablation studies in Section~\ref{sec:ablation}. Section~\ref{sec:threats} discusses potential threats to validity. Finally, Section~\ref{sec:conclusion} concludes the paper and outlines future research directions.

\section{Related Work}\label{sec:related}
\subsection{Cross-Project Learning for SRGMs}\label{subsec:cross}
A primary approach to mitigating data scarcity in software reliability modeling is leveraging knowledge from past projects. Early studies focused on reusing traditional SRGMs and their parameters from previous projects. Xie et al.~\cite{xie1997practical, xie1999software} first introduced the idea of directly adopting traditional SRGMs and their defect discovery rate parameters from similar past projects. However, they did not propose a method for selecting relevant past projects, and their experiments were conducted on a single dataset that had been manually pre-selected as similar. Rana et al.~\cite{rana2013evaluating} extended this concept to industrial automotive systems. However, they did not disclose the datasets used and, like Xie et al., did not provide a systematic project selection method.

Subsequent studies proposed heuristic or statistical criteria to identify relevant past projects. Honda et al.~\cite{honda2016case} introduced a classification-based approach, grouping projects based on lines of code, the number of test cases, and test density. Among these metrics, test density was found to be the most effective indicator for identifying similar past projects. Azturk et al.~\cite{azturk2021ensemble} proposed a statistical method using the Mann–Whitney U test to select projects with similar defect discovery distributions. While both approaches contributed to improving project selection, they relied on proprietary datasets that were not publicly available, which limited the reproducibility and generalizability of their findings.

Recent studies have explored deep learning-based cross-project transfer learning techniques that utilize defect discovery data from past projects. Hu et al.~\cite{hu2006early} proposed an ANN-based approach that learns defect discovery trends from past projects to predict software reliability in new ones. However, their experiments were conducted on a limited number of datasets and did not include a method for selecting relevant projects. San et al.~\cite{san2021deep, san2019dc} introduced Deep Cross-Project SRGM (DC-SRGM), a method that utilizes multiple past projects selected through clustering. Their approach applies k-means clustering based on the total number of defects, observation periods, and cross-correlation values to group similar projects. The selected datasets are then used to train an stacked LSTM model for reliability prediction. DC-SRGM is a representative deep learning-based method that employs multiple past projects, and it demonstrated improved prediction accuracy and robustness. However, it relied on proprietary datasets that were not publicly available, limiting reproducibility and external validation. Nagaraju et al.~\cite{nagaraju2022adaptive} proposed a method that employs the Granger causality test to identify the most relevant past project. An ANN model is trained using the selected dataset, and the method is evaluated on ten projects from Lyu’s benchmark dataset~\cite{lyu1996handbook}. While the method provides a structured approach to source project selection, it relies on a single-source dataset, which may limit predictive accuracy when the selected dataset is either small or insufficiently similar to the target.

\subsection{Availability of Past Project Datasets}\label{subsec:pastdatasets}
Although various approaches have been proposed for cross-project transfer learning in software reliability modeling, their effectiveness remains fundamentally limited by the availability and quality of past project datasets. The performance of these techniques depends heavily on the amount, diversity, and accessibility of such datasets. This subsection reviews studies that focus on providing publicly available software reliability datasets to support research and experimentation in this field.

The earliest datasets were introduced by Musa in 1979, comprising defect discovery data from 16 commercial off-the-shelf (COTS) software systems~\cite{musa1979software}. Each dataset is a Time Between Failures (TBF) sequence, where each entry represents the cumulative time between successive defect discoveries. Musa's datasets have been widely used in software reliability research over the past decades~\cite{costa2007exploring, costa2010genetic, nagaraju2017performance, li1993enhancing}. However, considering their age, these datasets may no longer reflect modern software development practices or defect discovery characteristics.

In 1996, Lyu compiled and expanded upon Musa’s datasets, providing a collection of 44 datasets as part of his handbook on software reliability engineering~\cite{lyu1996handbook}. This collection includes both TBF and Failure Count (FC) datasets, offering more diversity in modeling defect discovery trends. Lyu’s datasets remain one of the most frequently referenced sources in SRGM research. Nevertheless, these datasets originate from projects developed in the 1980s and 1990s, which limits their applicability to contemporary software systems.

More recently, Mi{\v{c}}ko et al.~\cite{mivcko2022applicability} published 88 TBF datasets collected from open-source software (OSS) projects. These datasets provide valuable insight into modern software reliability trends, particularly within open-source development environments. However, they do not include FC data, limiting their applicability in studies that require both data types. Additionally, they do not contain data from COTS software, which presents a gap in modeling the reliability of proprietary software systems.

The most extensive dataset compilation was conducted by Kim et al.~\cite{kim2024enhancing}, who collected 127 datasets from various prior studies. They integrated a wide range of datasets, including those by Musa and Lyu, which have been referenced in software reliability research for decades. While this collection represents the most comprehensive publicly available dataset compilation, it still faces limitations in terms of size, diversity, and potential bias.

Despite these efforts to provide past project datasets, publicly available software reliability datasets still exhibit several limitations. Many datasets are outdated, failing to reflect modern software development methodologies. Additionally, the number of available datasets remains limited, particularly for FC datasets, which are rarely provided in public repositories. Even among available datasets, many contain only a small number of data points, making it difficult to analyze long-term reliability trends. Furthermore, these datasets contain noise and missing or inconsistent records, making it hard to determine how many defects remained undetected at the end of data collection. These issues hinder the development and evaluation of traditional SRGMs and cross-project models.

\section{Proposed Methodology}\label{sec:method}

\begin{figure}[tbh]
    \centering
    \includegraphics[width=0.9\textwidth]{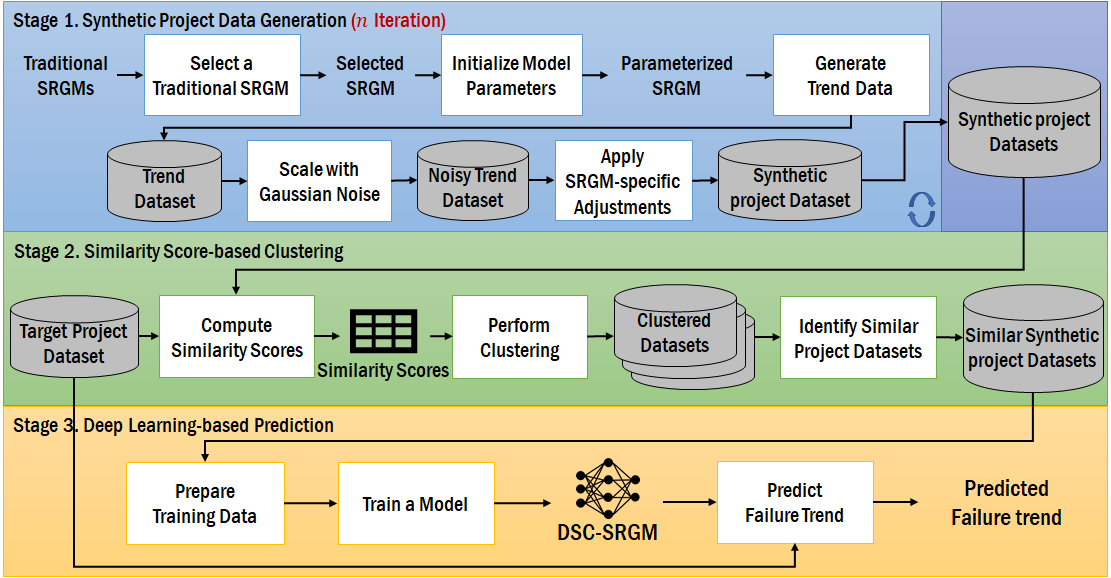}
    \caption{Overview of the DSC-SRGM framework. It consists of three stages: (1) Synthetic dataset generation using traditional SRGMs, (2) Similarity score-based clustering for relevant dataset selection, and (3) Deep learning-based reliability prediction.}
    \label{fig:overview}
\end{figure}

This study aims to enable accurate software reliability prediction even when both target and past project data are limited. To address this challenge, we propose Deep Synthetic Cross-project SRGM (DSC-SRGM), a novel approach that integrates synthetic data generation with cross-project transfer learning. Instead of relying on real-world datasets, DSC-SRGM synthesizes defect discovery data using well-established traditional SRGMs. The generated datasets retain the statistical characteristics of real-world defect discovery trends while offering greater flexibility in both volume and diversity. To identify datasets relevant to the target project, a similarity score-based clustering method using cross-correlation is applied. The selected datasets are then used to train a deep learning model, which enables the model to generalize effectively even in data-scarce environments. To the best of our knowledge, DSC-SRGM is the first approach to generate synthetic data using traditional SRGMs for cross-project transfer learning in software reliability modeling.

Figure~\ref{fig:overview} illustrates the overall architecture of the proposed DSC-SRGM framework, which consists of three main stages: synthetic data generation, similarity score-based clustering, and deep learning-based prediction. In the first stage, a traditional SRGM is randomly selected, and its parameters are initialized to simulate realistic defect discovery trends. Gaussian noise is added to the generated trends to increase variability, and adjustments are made to preserve key SRGM characteristics. In the second stage, similarity scores are computed between all pairs of datasets, including both synthetic datasets and the target dataset. Based on the resulting similarity matrix, clustering is performed to identify synthetic datasets that closely resemble the trend of the target project. In the third stage, a deep learning model is trained on the selected synthetic datasets to predict the future defect discovery trend in the target project. 

To complement the overview in Figure~\ref{fig:overview}, Algorithm~\ref{alg:dscsrgm} presents a step-by-step pseudocode of the proposed approach, offering a clear and reproducible description of the entire process.

\begin{algorithm}[tbh]
\caption{DSC-SRGM: Deep Synthetic Cross-Project Software Reliability Growth Modeling}
\label{alg:dscsrgm}
\begin{algorithmic}[1]
\Require Target dataset $D_{\text{target}}$, Number of synthetic datasets $n$, Number of clusters $K$
\Ensure Predicted defect discovery trend $\hat{D}_{\text{target}}$

\Statex
\Statex \textbf{Stage 1: Synthetic Data Generation}
\State Initialize synthetic dataset pool $\mathcal{D}_{\text{syn}} \gets \emptyset$
\For{$i = 1$ to $n$}
    \State Randomly select SRGM model $M_i \in \{$GO, YDSS, ISS, GG$\}$
    \State Sample parameters $\theta_i$ for $M_i$
    \State Generate cumulative trend $T_i(t) = M_i(t; \theta_i)$ until $T_i(t) \geq 0.95 \cdot a$ or $t = 512$
    \State Add Gaussian noise: $T_i(t) \gets T_i(t) \cdot (1 + \epsilon(t)), \epsilon(t) \sim \mathcal{N}(0, 0.001^2)$
    \State Apply SRGM-specific adjustment: $T_i(t) \gets \max(0, T_i(t), T_i(t{-}1))$
    \State Add $T_i(t)$ to $\mathcal{D}_{\text{syn}}$
\EndFor

\Statex
\Statex \textbf{Stage 2: Similarity Score-based Clustering}
\State Combine datasets: $\mathcal{D} \gets \mathcal{D}_{\text{syn}} \cup \{D_{\text{target}}\}$
\State Compute cross-correlation similarity matrix $S$ for all pairs in $\mathcal{D}$
\State Apply K-means clustering to $S$ with $K$ clusters
\State Identify cluster $\mathcal{C}_{\text{target}}$ containing $D_{\text{target}}$
\State Select synthetic datasets in $\mathcal{C}_{\text{target}}$ as training data $\mathcal{D}_{\text{train}}$

\Statex
\Statex \textbf{Stage 3: Deep Learning-based Prediction}
\State Convert $\mathcal{D}_{\text{train}}$ to training samples using a sliding window
\State Normalize each sample independently using Min-Max scaling
\State Train stacked LSTM model $f$ on normalized data
\State Initialize prediction input $X$ with most recent observed values in $D_{\text{target}}$
\State Initialize output sequence $\hat{D}_{\text{target}} \gets \emptyset$

\For{each prediction step}
    \State Apply Min-Max scaling to $X$: $X_{\text{norm}} \gets \text{MinMax}(X)$
    \State Predict next normalized value: $\hat{y}_{\text{norm}} \gets f(X_{\text{norm}})$
    \State Denormalize $\hat{y}_{\text{norm}}$ to obtain $\hat{y}$
    \State Append $\hat{y}$ to $\hat{D}_{\text{target}}$
    \State Update $X$ by appending $\hat{y}$ and removing the oldest value
\EndFor

\State \Return $\hat{D}_{\text{target}}$
\end{algorithmic}
\end{algorithm}

\subsection{Synthetic Project Data Generation}\label{subsec:synthetic}
The first stage in DSC-SRGM involves generating synthetic defect discovery datasets that resemble real-world software reliability trends. This process consists of the following steps:
\begin{enumerate}
    \item Selecting a traditional SRGM
    \item Initializing model parameters
    \item Generating trend data
    \item Scaling with Gaussian noise
    \item Applying SRGM-specific adjustments
\end{enumerate}

By repeating this procedure $n$ times, a collection of $n$ synthetic project datasets is obtained. These datasets serve as the foundation for cross-project model training in subsequent stages.

A common approach for synthetic data generation is to base it on existing real-world datasets~\cite{el2020practical}. However, defect discovery data from past projects are often scarce, unavailable, or proprietary, making such approaches impractical. Furthermore, synthetic datasets derived from real-world data tend to inherit biases specific to the original software project, which can limit their generalizability to other projects.

To address these limitations, DSC-SRGM synthesizes defect discovery datasets using well-established traditional SRGMs, which have been extensively studied and widely adopted in software reliability research. Traditional SRGMs offer a principled foundation for simulating representative defect discovery trends based on probabilistic formulations. The generation process incorporates parameter sampling, Gaussian noise injection, and SRGM-specific adjustments, allowing the creation of diverse and realistic defect discovery patterns. 

This model-based approach offers two key advantages. First, it enables synthetic data generation even in the complete absence of past project datasets, which is valuable in data-scarce environments. Second, it produces scalable and unbiased datasets that are not tied to any particular project. Importantly, this is the first study to generate synthetic project datasets solely using traditional SRGMs, without relying on any real-world defect data.

\subsubsection{Selecting a Traditional SRGM}\label{subsubsec:srgm}
The first step in synthetic dataset generation is to randomly select one of the traditional SRGMs listed in Table~\ref{tab:srgm_equations}. One model is selected with equal probability from four representative models: Goel-Okumoto (GO), Yamada Delayed S-shaped (YDSS), Inflection S-shaped (ISS), and Generalized Goel (GG). This uniform random selection ensures diversity in defect discovery trends across the generated datasets. Each synthetic defect discovery trend $T_i(t)$ is generated based on one of the traditional SRGMs defined by $m(t)$ in Table~\ref{tab:srgm_equations}.

\begin{table}[tbh]
\centering
\caption{Traditional SRGMs used in synthetic data generation.}
\label{tab:srgm_equations}
\scriptsize
\begin{tabular}{lccc}
\toprule
\textbf{Model} & \textbf{Type} & \textbf{Equation $m(t)$} & \textbf{Reference} \\
\midrule
Goel-Okumoto (GO) & Concave &
$m(t) = a \left(1 - e^{-bt}\right), \quad 0 < a,\; 0 < b < 1$ & \cite{goel1979time} \\
Yamada Delayed S-Shaped (YDSS) & S-shaped &
$m(t) = a \left(1 - (1 + bt)e^{-bt}\right), \quad 0 < a,\; 0 < b < 1$ & \cite{yamada2009s} \\
Inflection S-shaped (ISS) & S-shaped &
$m(t) = a \left( \frac{1 - e^{-bt}}{1 + \frac{1 - r}{r}e^{-bt}} \right), \quad 0 < a,\; 0 < b < 1,\; 0 < r < 1$ & \cite{ohba1984inflection} \\
Generalized Goel (GG) & Concave/S-shaped &
$m(t) = a \left(1 - e^{-bt}\right)^c, \quad 0 < a,\; 0 < b < 1,\; 0 < c$ & \cite{goel1982software} \\
\bottomrule
\end{tabular}
\end{table}

Each of these models represents a distinct defect discovery pattern. The GO model produces concave curves, while the YDSS and ISS models exhibit S-shaped curves. The GG model can represent both concave and S-shaped trends depending on its parameter settings. By encompassing both curve types, the selected SRGMs effectively represent the range of reliability trends observed in real-world software projects.

All models used in our approach are based on the perfect debugging assumption, where no new defects are introduced during the testing process. Each model includes the parameter $a$, which defines an upper bound on the cumulative number of detected defects. This property enables the generation process to control the length of each synthetic trend by specifying a target proportion of $a$, as described in the following section~\ref{subsubsec:trend}.

\subsubsection{Initializing Model Parameters}\label{subsubsec:parameters}
Once a traditional SRGM is selected, the next step is to initialize its model parameters. Each SRGM includes a set of parameters, denoted as $\theta$, that characterize the shape and scale of the defect discovery trend. For example, the parameter $a$ represents the total number of defects to be discovered, while $b$ controls the rate at which defects are detected. Additional shape parameters, such as $r$ and $c$, determine the curvature of the defect accumulation trend in the ISS and GG models, respectively.

Table~\ref{tab:srgm_parameters} presents the parameter initialization settings used in our synthetic data generation. Each parameter is sampled within the theoretical bounds defined by the SRGM equations in Table~\ref{tab:srgm_equations}. The total defect parameter $a$ is fixed at 100, as it affects only the vertical scale and does not influence the shape of the defect discovery trend. Since all sequences are later normalized, varying $a$ does not impact the learning process. The detection rate parameter $b$ is sampled from a log-uniform distribution over the range $(0.0001,\; 1.0]$, allowing the model to capture both rapid defect discovery over short periods and gradual discovery over longer durations. For the ISS model, the shape parameter $r$ is drawn from a uniform distribution in $(0.0001,\; 1.0]$. For the GG model, the shape parameter $c$ is sampled uniformly from $(0.01,\; 2.0]$. This range allows the model to represent both concave ($0 < c \leq 1$) and S-shaped ($1 < c < 2$) trends.

\begin{table}[tbh]
\centering
\caption{Parameter initialization ranges for traditional SRGMs.}
\label{tab:srgm_parameters}
\begin{tabular}{lll}
\toprule
\textbf{Model} & \textbf{Parameter} & \textbf{Range / Distribution} \\
\midrule
GO, YD, ISS, GG & $a$ (Total Failures) & $100$ (fixed) \\
GO, YD, ISS, GG & $b$ (Failure Detection Rate) & $(0.0001,\; 1.0]$, log-uniform \\
ISS & $r$ (Curve Shape Factor) & $(0.0001,\; 1.0]$, uniform \\
GG & $c$ (Curve Shape Factor) & $(0.01,\; 2.0]$, uniform \\
\bottomrule
\end{tabular}
\end{table}

\subsubsection{Generating Trend Data}\label{subsubsec:trend}
After the model parameters are initialized, a synthetic defect discovery trend is generated using the selected SRGM and its parameters. To ensure that the generated trend is realistic and suitable for model training, two constraints are applied. First, the maximum sequence length is limited to 512 time steps to prevent the generation of excessively long synthetic time series. Second, the generation is terminated once 95\% of the total predicted defects have been reached, i.e., when $T_i(t) \geq 0.95 \cdot a$, where $T_i(t)$ denotes the cumulative defect count at time $t$, and $a$ represents the total number of defects. This threshold, determined through ablation studies, helps avoid extended flat-tail regions in which no additional defects are detected. Such regions often arise near the end of the curve and may introduce bias into the learning process. These constraints allow the synthetic datasets to reflect realistic defect discovery trends while minimizing stagnation periods that degrade model performance.

\subsubsection{Scaling with Gaussian Noise}\label{subsubsec:noise}
To reflect real-world uncertainty, Gaussian noise is added to the generated defect discovery trend. This step prevents the synthetic data from being overly deterministic. The noise scaling process is defined as follows:

\begin{equation}
T_{\text{noisy}}(t) = T(t) \cdot (1 + \epsilon(t)), \quad \epsilon(t) \sim \mathcal{N}(0, 0.001^2)
\end{equation}

where $T(t)$ denotes the generated trend, and $\epsilon(t)$ follows a Gaussian distribution with a mean of $0$ and a standard deviation of $0.001$ (i.e., 0.1\%). This configuration was validated through ablation studies, as discussed in Section~\ref{sec:ablation}.

\subsubsection{Applying SRGM-Specific Adjustments}\label{subsubsec:adjustments}
Since defect discovery data represent cumulative failure counts, they must be non-negative and monotonically increasing. To enforce these properties, the following post-processing adjustment is applied:

\begin{equation}
T_{\text{final}}(t) = \max(0, T_{\text{noisy}}(t), T_{\text{final}}(t-1))
\end{equation}

where $T_{\text{final}}(t)$ denotes the adjusted cumulative defect count at time $t$. This adjustment ensures that the synthetic dataset maintains the essential characteristics of real-world defect discovery patterns.

\subsection{Similarity Score-based Clustering}\label{subsec:similarity}
After synthetic defect discovery datasets are generated, the next stage involves identifying the most relevant datasets for training the reliability prediction model. Since the generated datasets exhibit diverse defect discovery trends, it is essential to select only those that closely resemble the target project.

To achieve this, DSC-SRGM adopts a similarity score-based clustering approach using cross-correlation. Rather than relying on a single most similar dataset, DSC-SRGM selects a group of relevant datasets to improve model robustness and generalization. This strategy contrasts with the method proposed by Nagaraju et al.~\cite{nagaraju2022adaptive}, which selects only one source project based on the lowest p-value from the Granger causality test. Relying on a single dataset may lead to suboptimal performance, particularly when the selected dataset is small or fails to represent the defect discovery behavior of the target project. In contrast, clustering enables the selection of multiple similar datasets, improving both the diversity and representativeness of the training data.

Cross-correlation~\cite{egri2017cross} is adopted as the similarity metric due to its ability to compare sequences of variable lengths, which is a common characteristic of defect discovery data. San et al.~\cite{san2021deep} empirically demonstrated that cross-correlation outperforms DTW, a widely used similarity metric for time-series data, when applied to software reliability modeling. By using cross-correlation, DSC-SRGM enables more effective grouping of projects with similar defect discovery trends.

This stage comprises three steps: computing similarity scores, performing clustering, and identifying relevant synthetic datasets. The following subsections describe each step in detail.

\subsubsection{Computing Similarity Scores}\label{subsubsec:scores}
The first step is to compute pairwise similarity scores between all datasets, including synthetic datasets and the target project dataset. Cross-correlation, which is adopted in this study, measures similarity by shifting two datasets over various time lags and computing the correlation at each alignment. This method is particularly effective for comparing time series data that differ in length or exhibit temporal shifts~\cite{egri2017cross}. San et al.~\cite{san2021deep} empirically demonstrated that cross-correlation outperforms DTW, a widely used method for comparing variable-length time series data, when evaluating the similarity of defect discovery datasets in software reliability modeling. While their study also incorporates normalized total defect counts and observation periods into the similarity metric, we adopt cross-correlation alone to focus only on trend similarity. This decision prevents underestimation of similarity between datasets that share the same trend shape but differ in scale or duration.

Let \( x(t) \) and \( y(t) \) denote two defect discovery datasets. The cross-correlation (CC) at lag \( \tau \) is calculated as follows:

\begin{equation}
\text{CC}_{x,y}(\tau) =
\frac{
\sum_{t=1}^{n} \left( x(t) - \bar{x} \right) \left( y(t + \tau) - \bar{y} \right)
}{
\sqrt{
\sum_{t=1}^{n} \left( x(t) - \bar{x} \right)^2
}
\sqrt{
\sum_{t=1}^{n} \left( y(t + \tau) - \bar{y} \right)^2
}
}
\end{equation}

where \( \bar{x} \) and \( \bar{y} \) denote the means of \( x(t) \) and \( y(t) \), respectively. The value \( n \) represents the number of overlapping time points where both \( x(t) \) and \( y(t+\tau) \) are defined.

The similarity score between \( x(t) \) and \( y(t) \) is then defined as the maximum cross-correlation value computed across all possible time lags, including both positive and negative shifts:

\begin{equation}
\text{sim}(x, y) := \max_{\tau} \text{CC}_{x,y}(\tau)
\end{equation}

After computing similarity scores for all dataset pairs, a similarity matrix \( S \) is constructed, where each element \( S_{i,j} \) represents the similarity score between the \( i \)-th and \( j \)-th defect discovery datasets. This matrix is symmetric by definition and serves as the foundation for clustering similar datasets in the next step.

\subsubsection{Performing Clustering}\label{subsubsec:clustering}
Clustering is performed based on the similarity matrix to group datasets with similar defect discovery trends. K-means clustering is applied to partition the datasets into \( K \) clusters. Following the empirical findings of San et al.~\cite{san2021deep}, \( K = 3 \) is adopted, as their study demonstrated using the elbow method that three clusters provide effective grouping of software reliability datasets. This clustering step identifies subsets of datasets that exhibit similar temporal characteristics, which are then used to select relevant synthetic datasets in the subsequent step.

\subsubsection{Identifying Similar Project Data}\label{subsubsec:identifying}
After clustering, each project dataset is assigned to one of the \( K \) clusters. These clusters may contain both synthetic datasets and the target project dataset. To identify similar project datasets for training, we first locate the cluster that includes the target project dataset. All synthetic datasets within the same cluster are then selected and used for model training. This selection strategy ensures that the selected synthetic datasets used for training reflect defect discovery trends similar to the target project.

\subsection{Deep Learning-based Reliability Prediction}\label{subsec:learning}
In the final stage of DSC-SRGM, a deep learning model is trained using the synthetic datasets selected through similarity score-based clustering. The objective is to predict the future defect discovery trend of the target project based on its observed defect discovery data.

\subsubsection{Preparing Training Data}\label{subsubsec:preparing}
After selecting the relevant synthetic datasets, the data are preprocessed for training the deep learning model. First, the selected datasets are randomly split into training and validation sets at an 80:20 ratio. Then, a sliding window approach is applied to convert the data into a supervised learning format. Specifically, the cumulative defect counts from the past eight timesteps are used to predict the value at the next timestep. This formulation enables the model to learn temporal patterns in defect accumulation. Each training instance is then normalized independently using Min-Max scaling. Unlike conventional normalization applied at the dataset level, instance-wise normalization has been shown to improve generalization performance in time series modeling~\cite{kim2021reversible}. This approach mitigates the effect of scale variation across sequences and increases model stability during training.

\subsubsection{Training the Model}\label{subsubsec:training}
Once the training data are prepared, the prediction model is trained with a stacked LSTM architecture. The network consists of four LSTM layers, each with 128 hidden units. To prevent overfitting, a dropout rate of 0.2 is applied between layers. The model is trained for 300 epochs with a batch size of 64. Mean squared error (MSE) is used as the loss function, and the model with the lowest validation loss is selected for final prediction.

\subsubsection{Predicting Failure Trend}\label{subsubsec:predicting}
After training, the model is used to predict the future defect discovery trend of the target project based on the observed data. Since the model is designed for single-step forecasting, a recursive prediction strategy is employed. At each prediction step, the most recent input sequence is normalized using instance-wise Min-Max scaling, and the model predicts the next normalized value. This output is immediately denormalized to restore the predicted cumulative defect count. The denormalized prediction is then appended to the input sequence to form the next input. This process is repeated until the desired prediction length is reached.

\section{Experimental Setup}\label{sec:setup}
This section describes the experimental setup designed to evaluate the effectiveness of DSC-SRGM. The setup is organized according to four key research questions (RQ1–RQ4) and covers the datasets, baseline models, and evaluation metrics.

\subsection{Research Questions}\label{subsec:questions}
The experiments are designed to address the following research questions:

\begin{itemize}
    \item[] \textbf{RQ1.} Does DSC-SRGM outperform traditional SRGMs in terms of predictive accuracy?
    \item[] \textbf{RQ2.} Does training on synthetic datasets outperform training on real-world datasets in cross-project transfer learning?
    \item[] \textbf{RQ3.} Does combining synthetic and real-world datasets improve predictive performance compared to using only one type of dataset?
    \item[] \textbf{RQ4.} Does increasing the amount of synthetic data improve predictive performance?
\end{itemize}

\subsection{Datasets}\label{subsec:datasets}
We use a benchmark dataset collection compiled by Kim et al.~\cite{kim2024enhancing}, which aggregates various historical defect discovery datasets used in SRGM research. From this collection, we select 60 FC datasets with a sequence length of at least 15. This collection represents the largest dataset used in SRGM studies on knowledge reuse, encompassing diverse failure patterns and project characteristics. Table~\ref{tab:real_datasets} summarizes the basic information of the 60 defect datasets, including the number of failures, time span, and their original references. These datasets have been collected from a wide range of real-world software systems, covering both commercial and open-source projects~\cite{wood1996predicting, goel1982software, stringfellow2002empirical, huang2011estimation, wu2021study, satoh2000discrete, jeske2005adjusting, jeske2005some, kenny1993estimating, kaufman1999using, derriennic1995use, musa1987software, yang1996infinite, kaaniche1992discrete, ohba1982software, ohba1984software, xie1997practical, xie2007study, lyu1996handbook, martini1990software, misra1983software, garg2021decision, mullen1998lognormal, brooks1980analysis, li2011reliability, zhang2002calibrating, zhang2006software, tohma1989structural, tohma1989hyper, tohma1991parameter}.

\begin{table}[tbh]
\centering
\caption{Summary of 60 FC datasets employed in this study. All datasets were extracted from the benchmark dataset collection compiled by Kim et al.~\cite{kim2024enhancing}.}
\label{tab:real_datasets}
\scriptsize
\begin{tabular}{llll|llll|llll}
\toprule
\textbf{Dataset ID} & \textbf{Failures} & \textbf{Time} & \textbf{References} &
\textbf{Dataset ID} & \textbf{Failures} & \textbf{Time} & \textbf{References} &
\textbf{Dataset ID} & \textbf{Failures} & \textbf{Time} & \textbf{References} \\
\midrule
DS 1  & 100   & 20  & \cite{wood1996predicting} & DS 47 & 77   & 19  & \cite{musa1987software} & DS 92 & 421  & 72  & \cite{garg2021decision} \\
DS 2  & 120   & 19  & \cite{wood1996predicting} & DS 48 & 17   & 43  & \cite{musa1987software} & DS 94 & 360  & 73  & \cite{mullen1998lognormal} \\
DS 4  & 42    & 19  & \cite{wood1996predicting} & DS 50 & 136  & 21  & \cite{musa1987software} & DS 95 & 200  & 120 & \cite{mullen1998lognormal} \\
DS 8  & 2191  & 24  & \cite{goel1982software} & DS 57 & 4538 & 86  & \cite{yang1996infinite} & DS 97 & 1301 & 105 & \cite{brooks1980analysis} \\
DS 9  & 2621  & 24  & \cite{goel1982software} & DS 58 & 73   & 18  & \cite{kaaniche1992discrete} & DS 106& 74   & 43  & \cite{li2011reliability} \\
DS 10 & 4367  & 24  & \cite{goel1982software} & DS 60 & 23   & 15  & \cite{ohba1982software} & DS 107& 50   & 103 & \cite{li2011reliability} \\
DS 11 & 3937  & 24  & \cite{goel1982software} & DS 61 & 46   & 21  & \cite{ohba1984software} & DS 108& 58   & 164 & \cite{li2011reliability} \\
DS 12 & 176   & 18  & \cite{stringfellow2002empirical} & DS 63 & 328  & 19  & \cite{ohba1984software} & DS 109& 85   & 24  & \cite{li2011reliability} \\
DS 13 & 204   & 17  & \cite{stringfellow2002empirical} & DS 65 & 234  & 28  & \cite{xie1997practical} & DS 110& 54   & 24  & \cite{li2011reliability} \\
DS 15 & 167   & 21  & \cite{huang2011estimation} & DS 66 & 198  & 19  & \cite{xie1997practical} & DS 111& 54   & 46  & \cite{li2011reliability} \\
DS 16 & 8780  & 208 & \cite{wu2021study} & DS 67 & 144  & 17  & \cite{xie2007study} & DS 114& 1    & 62  & \cite{zhang2002calibrating} \\
DS 17 & 2251  & 193 & \cite{wu2021study} & DS 78 & 133  & 62  & \cite{lyu1996handbook} & DS 115& 1    & 40  & \cite{zhang2002calibrating} \\
DS 18 & 14608 & 188 & \cite{wu2021study} & DS 79 & 224  & 181 & \cite{lyu1996handbook} & DS 116& 1    & 21  & \cite{zhang2006software} \\
DS 19 & 5186  & 84  & \cite{satoh2000discrete} & DS 80 & 351  & 41  & \cite{lyu1996handbook} & DS 117& 1    & 21  & \cite{zhang2006software} \\
DS 25 & 181   & 34  & \cite{jeske2005adjusting, jeske2005some} & DS 81 & 188  & 144 & \cite{lyu1996handbook} & DS 121& 481  & 111 & \cite{tohma1989structural} \\
DS 26 & 22    & 17  & \cite{jeske2005adjusting, jeske2005some} & DS 82 & 367  & 73  & \cite{lyu1996handbook} & DS 122& 55   & 199 & \cite{tohma1989structural} \\
DS 27 & 111   & 36  & \cite{kenny1993estimating, kaufman1999using} & DS 83 & 100  & 140 & \cite{lyu1996handbook} & DS 123& 198  & 16  & \cite{tohma1989structural} \\
DS 28 & 357   & 24  & \cite{kenny1993estimating} & DS 85 & 211  & 32  & \cite{lyu1996handbook} & DS 124& 86   & 22  & \cite{tohma1989hyper} \\
DS 29 & 222   & 326 & \cite{derriennic1995use} & DS 89 & 461  & 81  & \cite{martini1990software} & DS 125& 266  & 46  & \cite{tohma1989hyper} \\
DS 30 & 79    & 36  & \cite{derriennic1995use} & DS 91 & 231  & 38  & \cite{misra1983software} & DS 126& 535  & 109 & \cite{tohma1991parameter} \\
\bottomrule
\end{tabular}
\end{table}

Each dataset is treated as a target dataset under a leave-one-out experimental setting. For each target, only the first 50\% of the dataset is assumed to be available for prediction. The model is then used to forecast the remaining 50\% of the cumulative defect discovery. This setup simulates early-stage reliability prediction scenarios, where defect data is inherently limited. The remaining 59 datasets, excluding the current target, are used as candidate training data for the real-data-based cross-project baseline. In this baseline, similar datasets are selected from the 59 real-world datasets. In contrast, DSC-SRGM does not use any real-world datasets from past projects. Instead, it generates 59 synthetic datasets using traditional SRGMs, matching the number of real-world datasets used in the baseline. This process is repeated for each of the 60 target projects to ensure a consistent and fair comparison across all experimental settings.

\subsection{Baseline Models}\label{subsec:baseline}

To evaluate the effectiveness of the proposed DSC-SRGM, we compare it with the following three baseline models:

\begin{itemize}
  \item[] \textbf{Best traditional SRGM:} For each target project, we fit six traditional models to the first 50\% of the corresponding dataset. The model with the lowest MSE is then selected as the best SRGM. The six models include: GO model~\cite{goel1979time}, YDSS model~\cite{yamada2009s}, ISS model~\cite{ohba1984inflection}, GG model~\cite{goel1982software}, Logistic (L) model~\cite{yamada1985software}, and Gompertz (G) model~\cite{yamada1985software}. Unlike the others, the L and G models are general-purpose statistical growth models rather than SRGMs specifically designed for software reliability. Nonetheless, they have been widely used in SRGM-related studies due to their flexible curve-fitting capabilities. Detailed definitions and selection results are described in Appendix~\ref{appendix:best_srgm}.
  \item[] \textbf{DC-SRGM (Real-only):} This baseline is adapted from the Deep Cross-Project SRGM (DC-SRGM) proposed by San et al.~\cite{san2021deep}, which is the most recent state-of-the-art model utilizing multiple relevant past projects. It is trained exclusively on real-world datasets selected via cross-correlation-based clustering. The model architecture and training settings are kept consistent with those of DSC-SRGM to ensure fair comparison.
  \item[] \textbf{Hybrid-SRGM:} This baseline combines both real-world and synthetic datasets for training. Similarity scores are computed on the combined set, and clustering is performed to select training data relevant to the target project. As with DC-SRGM and DSC-SRGM, it uses the same deep learning architecture and training settings. The only difference lies in the type of training data used.
\end{itemize}

\subsection{Evaluation Metrics}\label{subsec:metrics}
To evaluate predictive accuracy, we adopt three widely used regression metrics: Root Mean Squared Error (RMSE), Mean Absolute Error (MAE), and Mean Absolute Percentage Error (MAPE). These metrics quantify the difference between the predicted and actual cumulative defect values over the remaining 50\% of the target dataset.

The RMSE measures the square root of the average squared differences between predicted and actual values. It penalizes large errors more heavily and is formally defined as:
\begin{equation}
    \text{RMSE} = \sqrt{ \frac{1}{n} \sum_{i=1}^{n} \left( \hat{y}_i - y_i \right)^2 },
\end{equation}
where $y_i$ and $\hat{y}_i$ denote the actual and predicted values at time $i$, respectively, and $n$ is the number of prediction points.

The MAE computes the average of absolute differences between predicted and actual values. Unlike RMSE, it treats all errors equally without disproportionately penalizing larger deviations:
\begin{equation}
    \text{MAE} = \frac{1}{n} \sum_{i=1}^{n} \left| \hat{y}_i - y_i \right|.
\end{equation}

The MAPE expresses the prediction error relative to the actual values, allowing for scale-independent evaluation across datasets:
\begin{equation}
    \text{MAPE} = \frac{100}{n} \sum_{i=1}^{n} \left| \frac{\hat{y}_i - y_i}{y_i} \right|.
\end{equation}
Due to its normalized form, MAPE is particularly useful for comparing model performance across projects with varying failure scales.

\section{Experimental Results}\label{sec:results}
This section presents the experimental results for the four research questions (RQ1–RQ4). For each question, the proposed DSC-SRGM is evaluated against relevant baseline models using RMSE, MAE, and MAPE.

\subsection{RQ1: Does DSC-SRGM outperform traditional SRGMs in terms of predictive accuracy?}\label{subsec:rq1}
To address RQ1, we compare the predictive performance of DSC-SRGM with that of the best traditional SRGM for each target project. As described in Appendix~\ref{appendix:best_srgm}, the best traditional model is selected as the one that yields the lowest MSE when fitted to the first 50\% of each defect discovery dataset.

We first evaluate whether the performance differences are statistically significant. The Wilcoxon Signed-Rank Test~\cite{wilcoxon1992individual}, a non-parametric method for comparing two related models accross multiple paired samples, is applied to the RMSE, MAE, and MAPE values. The resulting $p$-values are 0.0118, 0.0145, and 0.0036, respectively. Since all values are below the 0.05 significance threshold, the differences are deemed statistically significant.

To assess overall predictive accuracy, we compute the median prediction errors across the 60 target datasets. As summarized in Table~\ref{tab:rq1_summary}, DSC-SRGM yields lower median errors than the best traditional SRGM across all three metrics. Specifically, DSC-SRGM achieves improvements of 13.7\%, 14.8\%, and 23.3\% in RMSE, MAE, and MAPE, respectively.

\begin{table}[tbh]
\centering
\caption{Median prediction errors and relative improvements of DSC-SRGM over the best traditional SRGM.}
\label{tab:rq1_summary}
\begin{tabular}{lccc}
\toprule
\textbf{Metric} & \textbf{Best Traditional SRGM} & \textbf{DSC-SRGM} & \textbf{Improvement} \\
\midrule
RMSE  & 16.8142 & \textbf{14.5060} & 13.7\% \\
MAE   & 14.3633 & \textbf{12.2295} & 14.8\% \\
MAPE  & 15.5474 & \textbf{11.9244} & 23.3\% \\
\bottomrule
\end{tabular}
\end{table}

\begin{figure}[tbh]
    \centering
    \includegraphics[width=\textwidth]{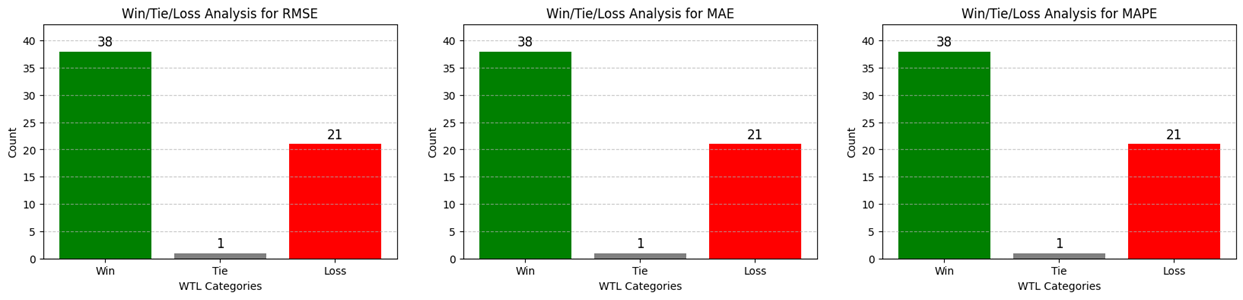}
    \caption{Win/Tie/Loss analysis between DSC-SRGM and the best traditional SRGM across 60 project datasets.}
    \label{fig:rq1_wtl}
\end{figure}

To further analyze model performance at the individual dataset level, we conduct a Win/Tie/Loss (WTL) analysis. A win is counted if DSC-SRGM's prediction error is at least 5\% lower than that of the baseline. A tie is recorded when the relative difference is within 5\%, and a loss otherwise. Figure~\ref{fig:rq1_wtl} presents a bar chart of the WTL outcomes for each metric. Across all three metrics, DSC-SRGM wins in 38 out of 60 project datasets, ties in 1 case, and loses in 21 cases. These results indicate that DSC-SRGM consistently outperforms the baseline across the majority of datasets.

\begin{figure}[tbh]
    \centering
    \includegraphics[width=\textwidth]{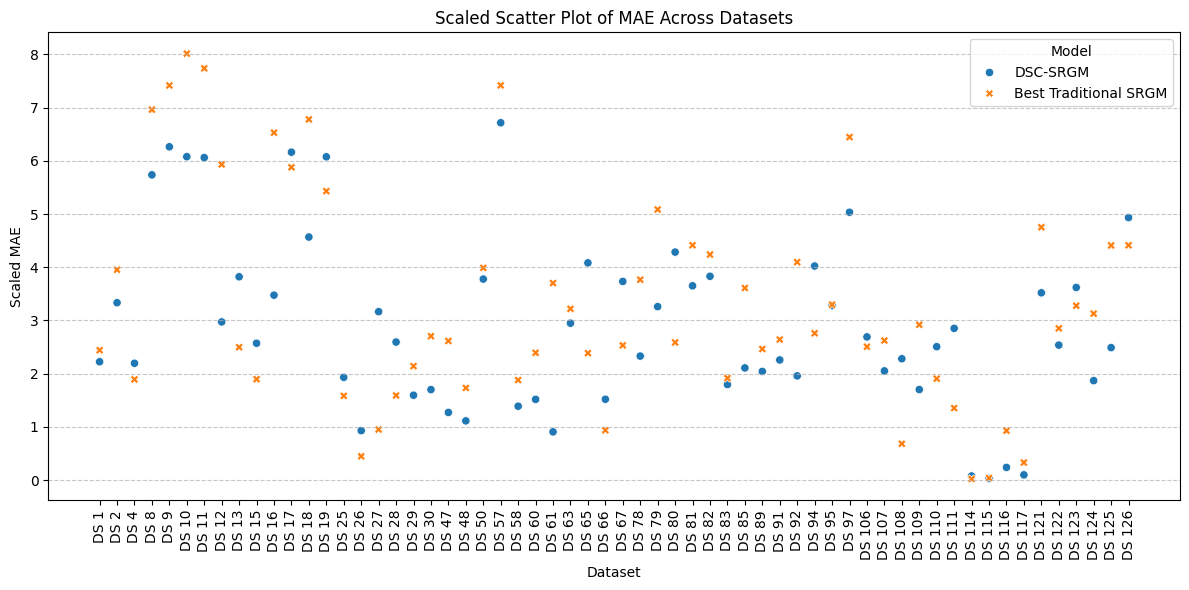}
    \caption{Scatter plot of log-scaled MAE values for each dataset. Blue circles: DSC-SRGM; Orange crosses: Best Traditional SRGM.}
    \label{fig:rq1_scatter}
\end{figure}

To complement this analysis, we visualize the prediction results using a scatter plot of log-scaled MAE values for each dataset. A logarithmic transformation is applied to the MAE values to reduce the influence of large-magnitude errors and improve interpretability. Figure~\ref{fig:rq1_scatter} presents the results, where blue circles represent DSC-SRGM and orange crosses denote the best traditional SRGM. In most cases, the blue circles lie below the orange crosses, indicating that DSC-SRGM yields lower MAE values for the majority of datasets. This visual evidence further supports that DSC-SRGM outperforms traditional SRGMs in predicting defect discovery trends. Similar scatter plots for RMSE and MAPE are provided in Appendix~\ref{appendix:scatter}, showing consistent trends with those observed for MAE.

\subsection{RQ2: Does training models on synthetic datasets outperform those trained on real-world datasets in cross-project transfer learning?}
\label{subsec:rq2}
To address RQ2, we compare the proposed DSC-SRGM, which is trained on synthetic datasets, with DC-SRGM, a cross-project baseline trained on real-world datasets. Both models share the same architecture and training configurations, differing only in the source of training data. While DC-SRGM selects training datasets from past projects using similarity-based clustering, DSC-SRGM applies the same strategy to synthetically generated datasets.

To determine whether the observed performance differences are statistically significant, we apply the Wilcoxon Signed-Rank Test. The resulting $p$-values for RMSE, MAE, and MAPE are 0.0157, 0.0181, and 0.0315, respectively. Since all values fall below the 0.05 significance threshold, we conclude that the performance differences are statistically significant.

We then compare the overall predictive performance based on the median errors across all 60 target datasets. As summarized in Table~\ref{tab:rq2_summary}, DSC-SRGM achieves lower median errors than DC-SRGM across all three metrics. Specifically, the proposed method reduces RMSE by 32.1\%, MAE by 32.2\%, and MAPE by 5.4\%. These findings suggest that synthetic datasets can be a viable and effective alternative to real-world datasets in cross-project transfer learning for software reliability prediction.

\begin{table}[tbh]
\centering
\caption{Median prediction errors and relative improvements of DSC-SRGM compared to DC-SRGM.}
\label{tab:rq2_summary}
\begin{tabular}{lccc}
\toprule
\textbf{Metric} & \textbf{DC-SRGM} & \textbf{DSC-SRGM} & \textbf{Improvement} \\
\midrule
RMSE  & 21.3515 & \textbf{14.5060} & 32.1\% \\
MAE   & 18.0308 & \textbf{12.2295} & 32.2\% \\
MAPE  & 12.5987 & \textbf{11.9244} & 5.4\%  \\
\bottomrule
\end{tabular}
\end{table}

\begin{figure}[tbh]
    \centering
    \includegraphics[width=\textwidth]{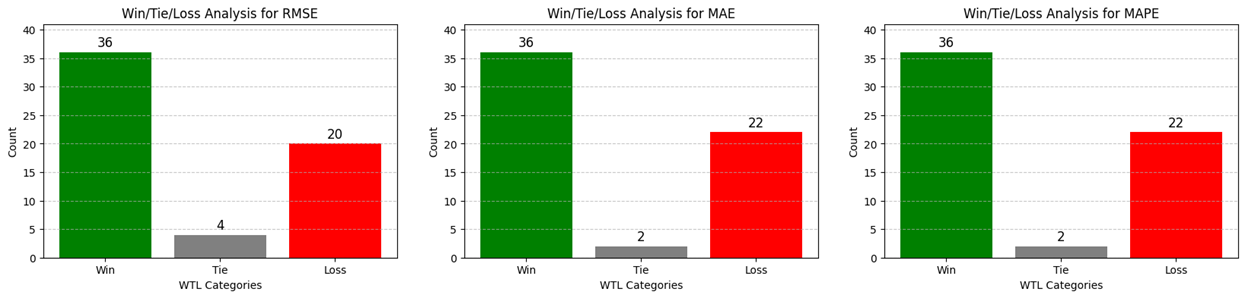}
    \caption{Win/Tie/Loss analysis comparing DSC-SRGM to DC-SRGM across 60 datasets.}
    \label{fig:rq2_wtl}
\end{figure}

To further evaluate performance at the individual dataset level, we conduct a WTL analysis. Figure~\ref{fig:rq2_wtl} presents the WTL outcomes across all 60 datasets for each evaluation metric. DSC-SRGM wins in 36 out of 60 datasets across all three metrics, while ties occur in only 2 to 4 cases depending on the metric. These results indicate that DSC-SRGM provides more consistent and robust performance than the baseline trained on real-world data.

\begin{figure}[tbh]
    \centering
    \includegraphics[width=\textwidth]{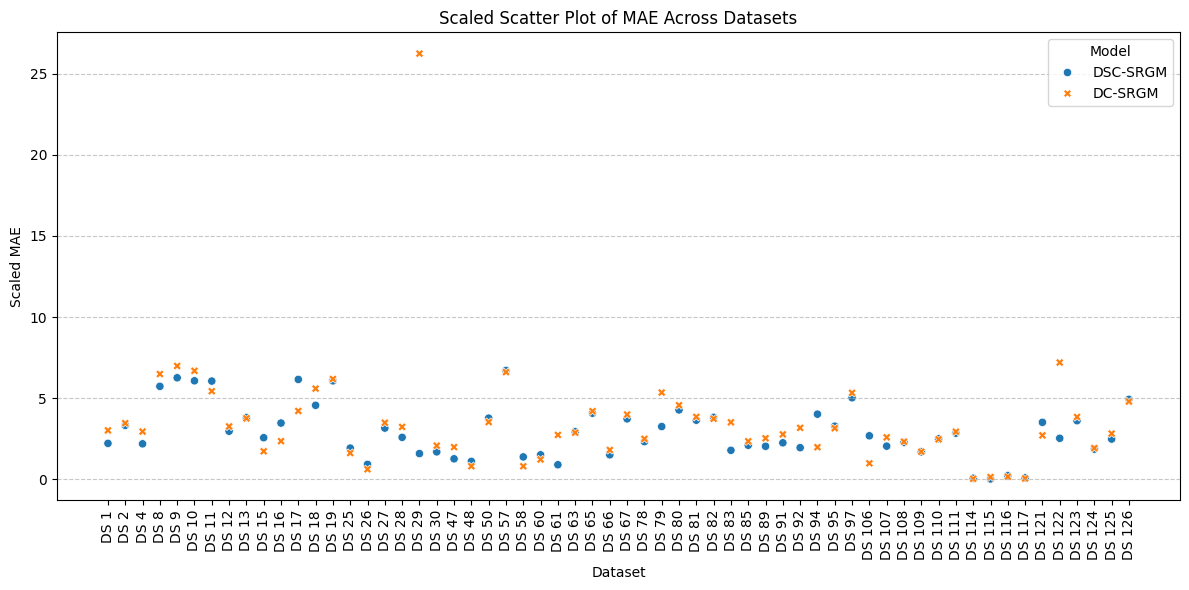}
    \caption{Scatter plot of log-scaled MAE values across datasets. Blue: DSC-SRGM, Orange: DC-SRGM.}
    \label{fig:rq2_scatter}
\end{figure}

Figure~\ref{fig:rq2_scatter} visualizes the predictive performance for each dataset using a scatter plot of log-scaled MAE values. Blue circles represent DSC-SRGM, and orange crosses denote DC-SRGM. In most cases, the blue markers lie below the orange ones, indicating that DSC-SRGM achieves lower MAE values. Furthermore, DC-SRGM exhibits several extreme prediction errors, whereas DSC-SRGM maintains stable and bounded performance. These results demonstrate that synthetic-data-based learning is not only more accurate on average but also more robust to outliers, thereby leading to more reliable predictions in practice.

In summary, these findings support the conclusion that DSC-SRGM, even without access to real-world data, can outperform state-of-the-art cross-project transfer learning methods. Carefully generated and properly selected synthetic datasets can be a scalable and effective alternative for software reliability prediction. Additional scatter plots based on RMSE and MAPE are provided in Appendix~\ref{appendix:scatter}.

\subsection{RQ3: Does combining synthetic and real-world datasets improve predictive performance compared to using only one type of dataset?}
\label{subsec5-3}

To answer RQ3, we compare three training strategies: DSC-SRGM trained exclusively on synthetic datasets, DC-SRGM trained solely on real-world datasets, and Hybrid-SRGM trained on a combination of both.

To evaluate whether the performance differences among the three models are statistically significant, we apply the Friedman test~\cite{friedman1937use}, a non-parametric test appropriate for comparing more than two related models. A $p$-value below 0.05 indicates that at least one model performs significantly differently from the others. The Friedman test yields $p$-values of $2.4847 \times 10^{-10}$ (RMSE), $2.4625 \times 10^{-10}$ (MAE), and $1.0787 \times 10^{-6}$ (MAPE), confirming that the observed differences are statistically significant.

Given these results, we further perform pairwise comparisons using the Wilcoxon signed-rank test. Table~\ref{tab:rq3_wilcoxon} reports the $p$-values for all pairwise comparisons across the three evaluation metrics.

\begin{table}[tbh]
\centering
\caption{Wilcoxon signed-rank test $p$-values for pairwise model comparisons in RQ3.}
\label{tab:rq3_wilcoxon}
\begin{tabular}{lccc}
\toprule
\textbf{Model Pair} & \textbf{RMSE} & \textbf{MAE} & \textbf{MAPE} \\
\midrule
DC-SRGM vs. Hybrid-SRGM & 0.6428 & 0.5658 & 0.9296 \\ 
DSC-SRGM vs. Hybrid-SRGM & \textbf{0.0097} & \textbf{0.0104} & \textbf{0.0113} \\
\bottomrule
\end{tabular}
\end{table}

The results confirm that the performance differences between DSC-SRGM and Hybrid-SRGM are statistically significant across all three evaluation metrics. In contrast, no statistically significant difference is observed between Hybrid-SRGM and DC-SRGM. These findings suggest that the impact of synthetic data depends on its integration strategy.

Table~\ref{tab:rq3_median} presents the median prediction errors across the 60 target datasets. DSC-SRGM achieves the lowest median values for all three metrics. In contrast, Hybrid-SRGM performs worse than both DSC-SRGM and DC-SRGM, indicating that naively combining real and synthetic datasets may introduce factors that negatively impact prediction accuracy.

\begin{table}[tbh]
\centering
\caption{Median prediction errors across 60 target datasets.}
\label{tab:rq3_median}
\begin{tabular}{lccc}
\toprule
\textbf{Model} & \textbf{RMSE} & \textbf{MAE} & \textbf{MAPE} \\
\midrule
DC-SRGM (Real-only)   & 21.3515 & 18.0308 & 12.5987 \\
DSC-SRGM (Synthetic-only) & \textbf{14.5060} & \textbf{12.2295} & \textbf{11.9244} \\
Hybrid-SRGM (Real+Synthetic) & 22.5195 & 20.0049 & 14.2467 \\
\bottomrule
\end{tabular}
\end{table}

Next, we perform a WTL analysis to examine model performance at the individual dataset level. Table~\ref{tab:rq3_wtl} presents the results for all pairwise comparisons. DSC-SRGM consistently outperforms both DC-SRGM and Hybrid-SRGM across the majority of datasets. In contrast, Hybrid-SRGM does not show a consistent advantage over DC-SRGM, suggesting that combining real and synthetic data does not guarantee improved performance.

\begin{table}[tbh]
\centering
\caption{Win/Tie/Loss analysis for pairwise comparisons among the three learning strategies.}
\label{tab:rq3_wtl}
\begin{tabular}{lcccc}
\toprule
\textbf{Comparison} & \textbf{Metric} & \textbf{Win} & \textbf{Tie} & \textbf{Loss} \\
\midrule
DSC-SRGM vs. DC-SRGM     & RMSE & \textbf{36} & 4 & 20 \\
               & MAE  & \textbf{36} & 2 & 22 \\
               & MAPE & \textbf{36} & 2 & 22 \\
\midrule
DSC-SRGM vs. Hybrid-SRGM & RMSE & \textbf{38} & 4 & 18 \\
               & MAE  & \textbf{35} & 8 & 17 \\
               & MAPE & \textbf{35} & 8 & 17 \\
\midrule
DC-SRGM vs. Hybrid-SRGM  & RMSE & \textbf{29} & 6 & 25 \\
               & MAE  & \textbf{31} & 6 & 23 \\
               & MAPE & \textbf{31} & 5 & 24 \\
\bottomrule
\end{tabular}
\end{table}

\begin{figure}[tbh]
    \centering
    \includegraphics[width=\textwidth]{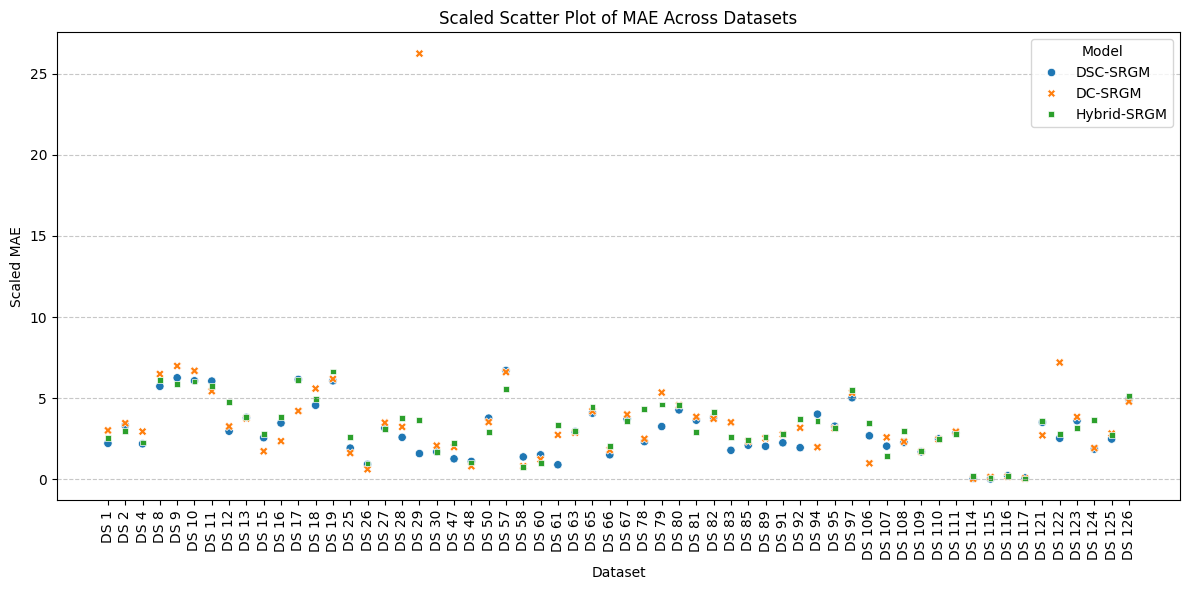}
    \caption{Scatter plot of log-scaled MAE values for three models across datasets.}
    \label{fig:rq3_scatter}
\end{figure}

Figure~\ref{fig:rq3_scatter} visualizes the log-scaled MAE values across all datasets to facilitate visual comparison of the models. DSC-SRGM (blue circles) maintains low and stable errors across most datasets. Although Hybrid-SRGM (green squares) occasionally mitigates extreme outliers, it generally fails to outperform either DSC-SRGM or DC-SRGM.

In summary, the results show that DSC-SRGM outperforms both DC-SRGM and Hybrid-SRGM in terms of predictive accuracy and consistency. While Hybrid-SRGM does not demonstrate overall performance improvements, it helps reduce extreme prediction errors occasionally observed in the real-data-based DC-SRGM. This suggests that incorporating synthetic data can improve the robustness of the model in certain cases.

\subsection{RQ4: Does increasing the amount of synthetic data improve Predictive Performance?}
RQ4 investigates the effect of synthetic data quantity on the prediction performance of DSC-SRGM. Using the default setting of $n = 59$ synthetic datasets, corresponding to the number of available real-world reference datasets, as a baseline, we evaluate model performance as the number of synthetic datasets increases to $2n$, $3n$, $4n$, and $5n$.

To assess whether these performance differences are statistically significant, we apply the Friedman test. The resulting $p$-values are $1.0597 \times 10^{-10}$ (RMSE), $8.9624 \times 10^{-21}$ (MAE), and $7.8442 \times 10^{-16}$ (MAPE), all well below the 0.05 significance threshold. These results confirm that the quantity of synthetic data has a statistically significant impact on predictive performance.

Table~\ref{tab:rq4_performance} reports the median RMSE, MAE, and MAPE values across all target datasets. The best performance is observed at $n = 59$ for RMSE and MAE, and at $3n$ for MAPE. To further assess the performance consistency, we conduct a WTL analysis comparing the baseline ($n = 59$) with each of the extended synthetic data configurations. As shown in Table~\ref{tab:rq4_wtl}, the baseline consistently achieves the most wins across all metrics. Notably, the number of wins increases as the size of the synthetic dataset increases, suggesting that using excessive synthetic data may degrade model accuracy.

\begin{table}[tbh]
\centering
\caption{Median prediction errors of DSC-SRGM under varying synthetic dataset sizes. The best value for each metric is highlighted in bold.}
\label{tab:rq4_performance}
\begin{tabular}{lccc}
\toprule
\textbf{Model} & \textbf{Median RMSE} & \textbf{Median MAE} & \textbf{Median MAPE} \\
\midrule
DSC-SRGM (Synthetic Data $n$)   & \textbf{14.5060} & \textbf{12.2295} & 11.9244 \\
DSC-SRGM (Synthetic Data $2n$)  & 18.8038 & 16.6539 & 10.9609 \\
DSC-SRGM (Synthetic Data $3n$)  & 18.8730 & 17.1317 & \textbf{9.8350} \\
DSC-SRGM (Synthetic Data $4n$)  & 22.8615 & 18.5824 & 11.9356 \\
DSC-SRGM (Synthetic Data $5n$)  & 19.3294 & 16.6462 & 12.6322 \\
\bottomrule
\end{tabular}
\end{table}

\begin{table}[tbh]
\centering
\caption{Win/Tie/Loss analysis comparing DSC-SRGM ($n$) with larger synthetic dataset configurations.}
\label{tab:rq4_wtl}
\begin{tabular}{lccc|ccc|ccc}
\toprule
\multirow{2}{*}{\textbf{Comparison}} & \multicolumn{3}{c|}{\textbf{RMSE}} & \multicolumn{3}{c|}{\textbf{MAE}} & \multicolumn{3}{c}{\textbf{MAPE}} \\
 & Win & Tie & Loss & Win & Tie & Loss & Win & Tie & Loss \\
\midrule
(Synthetic Data $n$) vs. (Synthetic Data $2n$) & \textbf{28} & 8 & 24 & \textbf{29} & 7 & 24 & \textbf{29} & 7 & 24 \\
(Synthetic Data $n$) vs. (Synthetic Data $3n$) & \textbf{27} & 9 & 24 & \textbf{27} & 9 & 24 & \textbf{26} & 10 & 24 \\
(Synthetic Data $n$) vs. (Synthetic Data $4n$) & \textbf{33} & 5 & 22 & \textbf{34} & 5 & 21 & \textbf{32} & 7 & 21 \\
(Synthetic Data $n$) vs. (Synthetic Data $5n$) & \textbf{32} & 6 & 22 & \textbf{32} & 5 & 23 & \textbf{32} & 6 & 22 \\
\bottomrule
\end{tabular}
\end{table}

These findings suggest that while an appropriate amount of synthetic data can improve predictive performance, excessively increasing the quantity may instead degrade model accuracy. This performance degradation can be interpreted from three perspectives. First, increasing the number of synthetic datasets expands the cluster used for training. This expansion may incorporate data that are less similar to the target project, thereby reducing the overall relevance and quality of the training data. Second, DSC-SRGM operates under a fixed input sequence length. When the diversity of training data increases, it becomes more difficult for the model to focus on meaningful patterns within each input. This can hinder its ability to generalize effectively. Third, large-scale synthetic data generation may introduce redundant or noisy sequences. These less informative or inconsistent patterns can confuse the learning process and negatively affect prediction accuracy.

In summary, while synthetic data is effective for cross-project software reliability prediction, its benefit depends on maintaining a suitable data quantity and preserving high similarity to the target project.

\section{Ablation Studies}
\label{sec:ablation}

Beyond the four primary research questions, we conduct ablation studies to investigate the impact of key components in the synthetic data generation process. Specifically, we evaluate (1) the termination threshold used in synthetic dataset generation and (2) the level of Gaussian noise applied to the generated data.

\subsection{Impact of Synthetic Data Length Threshold}\label{subsec:length}

\begin{figure}[tbh]
    \centering
    \includegraphics[width=0.5\textwidth]{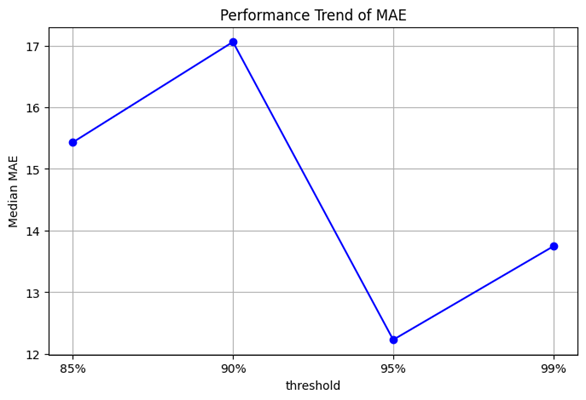}
    \caption{Impact of termination threshold on prediction performance (MAE).}
    \label{fig:threshold_ablation}
\end{figure}

To determine the optimal point to stop the generation of each synthetic dataset, we vary the termination threshold as a percentage of the total cumulative defect count. The thresholds tested include 85\%, 90\%, 95\%, and 99\%. As shown in Figure~\ref{fig:threshold_ablation}, the model trained on synthetic datasets generated up to the 95\% threshold achieves the lowest median MAE across all target datasets.

This result suggests that terminating the generation of each synthetic dataset at 95\% of the total defect count achieves an effective balance between informative patterns and noise suppression. Datasets generated up to this point exclude flat-tail regions that offer limited learning value, while preserving sufficient temporal variation for effective model training.

\subsection{Impact of Noise Level}

\begin{figure}[tbh]
    \centering
    \includegraphics[width=0.5\textwidth]{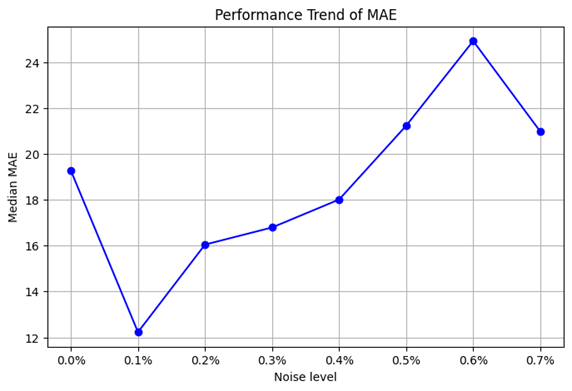}
    \caption{Effect of Gaussian noise level on prediction performance (MAE).}
    \label{fig:noise_ablation}
\end{figure}

We further investigate the impact of noise by injecting Gaussian noise into each synthetic dataset. The tested noise levels range from 0.0\% to 0.7\%, with increments of 0.1\%. Figure~\ref{fig:noise_ablation} shows the resulting trend in median MAE across all target datasets.

The best performance is observed at a noise level of 0.1\%, indicating that a small amount of noise improves generalization by increasing data variability without significantly distorting the underlying defect discovery patterns. In contrast, both the absence of noise and the injection of excessive noise lead to performance degradation. This is likely due to insufficient data diversity or overly distorted training inputs.

\section{Threats to Validity}\label{sec:threats}

\textbf{Internal Validity} Several design choices, including parameter ranges for synthetic data generation, noise levels, and clustering configurations, may influence the prediction performance. Although ablation studies were conducted to determine appropriate values, there remains the possibility that different parameter settings may lead to improved or degraded results. Additionally, the effectiveness of clustering depends on the accuracy of similarity scores, which are computed using cross-correlation. This method may not fully capture the trend similarity between datasets in all cases.

\noindent \textbf{External Validity} Although the proposed model was evaluated using 60 real-world defect discovery datasets, these datasets primarily originate from publicly available sources and may not fully reflect the characteristics of modern, large-scale, or safety-critical software systems. Consequently, the generalizability of the results to industrial contexts with different development practices may be limited. Additional validation using contemporary proprietary datasets is necessary to confirm broader applicability.

\noindent \textbf{Construct Validity} This study employs common regression metrics (RMSE, MAE, MAPE) to evaluate prediction performance. While these metrics are widely used, these metrics may not fully capture domain-specific aspects of software reliability. In addition, the fixed 50\% observation window may not suit all deployment scenarios, where data availability varies by development stage or context. Alternative evaluation criteria could provide a more comprehensive and context-aware assessment of model effectiveness.

\section{conclusion}\label{sec:conclusion}
This study proposed DSC-SRGM, a novel deep learning-based software reliability growth model that integrates synthetic data generation with cross-project transfer learning. The method leverages traditional SRGMs to generate diverse and statistically meaningful synthetic datasets and applies cross-correlation-based clustering to select datasets that resemble the target project. This approach enables accurate and robust reiliability modeling, particularly in data-scarce environments.

To the best of our knowledge, this is the first work that applies traditional SRGMs to generate synthetic data for cross-project transfer learning in software reliability prediction. By removing the dependency on real-world past project data, DSC-SRGM enables predictive modeling even when such data are unavailable or insufficient.

Comprehensive experiments on 60 real-world datasets demonstrate that DSC-SRGM consistently outperforms both traditional SRGMs and deep learning-based transfer learning models trained on real data. Notably, models trained solely on synthetic data achieve not only higher predictive accuracy but also greater stability across diverse projects. However, the results also show that excessive use of synthetic data may degrade performance, underscoring the importance of controlling both the quantity and similarity of training datasets.

These findings suggest that synthetic data can serve as an effective and scalable alternative to proprietary project data in software reliability prediction. DSC-SRGM is particularly well suited for early-stage testing and safety-critical domains, where defect discovery data are inherently limited.

Future work will explore improving synthetic data quality through advanced sampling strategies and generative constraints, refining clustering methods to better preserve project relevance, and adopting more flexible input structures to capture complex defect discovery patterns.

\bibliographystyle{unsrt}  
\bibliography{references} 

\begin{thebibliography}{10}

\bibitem{wood1996software}
Alan Wood.
\newblock Software reliability growth models.
\newblock {\em Tandem technical report}, 96(130056):900, 1996.

\bibitem{okumoto1979optimum}
Kazu Okumoto and Amrit~L Goel.
\newblock Optimum release time for software systems based on reliability and cost criteria.
\newblock {\em Journal of Systems and Software}, 1:315--318, 1979.

\bibitem{rani2021entropy}
Pooja Rani and GS~Mahapatra.
\newblock Entropy based enhanced particle swarm optimization on multi-objective software reliability modelling for optimal testing resources allocation.
\newblock {\em Software Testing, Verification and Reliability}, 31(6):e1765, 2021.

\bibitem{shooman1972probabilistic}
Martin~L Shooman.
\newblock Probabilistic models for software reliability prediction.
\newblock In {\em Statistical computer performance evaluation}, pages 485--502. Elsevier, 1972.

\bibitem{goel1985software}
Amrit~L. Goel.
\newblock Software reliability models: Assumptions, limitations, and applicability.
\newblock {\em IEEE Transactions on software engineering}, (12):1411--1423, 1985.

\bibitem{goel1979software}
Amrit~L Goel.
\newblock Software error detection model with applications.
\newblock {\em Journal of Systems and Software}, 1:243--249, 1979.

\bibitem{goel1979time}
Amrit~L Goel and Kazu Okumoto.
\newblock Time-dependent error-detection rate model for software reliability and other performance measures.
\newblock {\em IEEE transactions on Reliability}, 28(3):206--211, 1979.

\bibitem{yamada2009s}
Shigeru Yamada, Mitsuru Ohba, and Shunji Osaki.
\newblock S-shaped reliability growth modeling for software error detection.
\newblock {\em IEEE Transactions on reliability}, 32(5):475--484, 2009.

\bibitem{ohba1984inflection}
Mitsuru Ohba.
\newblock Inflection s-shaped software reliability growth model.
\newblock In {\em Stochastic Models in Reliability Theory: Proceedings of a Symposium Held in Nagoya, Japan, April 23--24, 1984}, pages 144--162. Springer, 1984.

\bibitem{goel1982software}
Amrit~L Goel.
\newblock {\em Software reliability modelling and estimation techniques}.
\newblock Rome Air Development Center, Air Force Systems Command, 1982.

\bibitem{kapur2011software}
PK~Kapur, Hoang Pham, Anshu Gupta, PC~Jha, et~al.
\newblock {\em Software reliability assessment with OR applications}, volume 364.
\newblock Springer, 2011.

\bibitem{sharma2010selection}
Kapil Sharma, Rakesh Garg, Chander~Kumar Nagpal, and Ramesh~Kumar Garg.
\newblock Selection of optimal software reliability growth models using a distance based approach.
\newblock {\em IEEE Transactions on Reliability}, 59(2):266--276, 2010.

\bibitem{stringfellow2002empirical}
Catherine Stringfellow and A~Amschler Andrews.
\newblock An empirical method for selecting software reliability growth models.
\newblock {\em Empirical Software Engineering}, 7:319--343, 2002.

\bibitem{jabeen2019improved}
Gul Jabeen, Ping Luo, and Wasif Afzal.
\newblock An improved software reliability prediction model by using high precision error iterative analysis method.
\newblock {\em Software Testing, Verification and Reliability}, 29(6-7):e1710, 2019.

\bibitem{li1993enhancing}
Naixin Li and Yashwant~K Malaiya.
\newblock Enhancing accuracy of software reliability prediction.
\newblock In {\em Proceedings of 1993 IEEE International Symposium on Software Reliability Engineering}, pages 71--79. IEEE, 1993.

\bibitem{cinque2017debugging}
Marcello Cinque, Domenico Cotroneo, Antonio Pecchia, Roberto Pietrantuono, and Stefano Russo.
\newblock Debugging-workflow-aware software reliability growth analysis.
\newblock {\em Software Testing, Verification and Reliability}, 27(7):e1638, 2017.

\bibitem{yang2010generic}
Bo~Yang, Xiang Li, Min Xie, and Feng Tan.
\newblock A generic data-driven software reliability model with model mining technique.
\newblock {\em Reliability Engineering \& System Safety}, 95(6):671--678, 2010.

\bibitem{pai2006software}
Ping-Feng Pai and Wei-Chiang Hong.
\newblock Software reliability forecasting by support vector machines with simulated annealing algorithms.
\newblock {\em Journal of Systems and Software}, 79(6):747--755, 2006.

\bibitem{das2011failure}
M{\'a}rcio das Chagas~Moura, Enrico Zio, Isis~Didier Lins, and Enrique Droguett.
\newblock Failure and reliability prediction by support vector machines regression of time series data.
\newblock {\em Reliability Engineering \& System Safety}, 96(11):1527--1534, 2011.

\bibitem{karunanithi1992prediction}
Nachimuthu Karunanithi, Darrell Whitley, and Yashwant~K. Malaiya.
\newblock Prediction of software reliability using connectionist models.
\newblock {\em IEEE Transactions on software engineering}, 18(7):563, 1992.

\bibitem{cai2001neural}
Kai-Yuan Cai, Lin Cai, Wei-Dong Wang, Zhou-Yi Yu, and David Zhang.
\newblock On the neural network approach in software reliability modeling.
\newblock {\em Journal of Systems and Software}, 58(1):47--62, 2001.

\bibitem{wang2018software}
Jinyong Wang and Ce~Zhang.
\newblock Software reliability prediction using a deep learning model based on the rnn encoder--decoder.
\newblock {\em Reliability Engineering \& System Safety}, 170:73--82, 2018.

\bibitem{behera2018software}
Ranjan~Kumar Behera, Suyash Shukla, Santanu~Kumar Rath, and Sanjay Misra.
\newblock Software reliability assessment using machine learning technique.
\newblock In {\em Computational Science and Its Applications--ICCSA 2018: 18th International Conference, Melbourne, VIC, Australia, July 2-5, 2018, Proceedings, Part V 18}, pages 403--411. Springer, 2018.

\bibitem{gusmanov2019cnn}
Kamill Gusmanov.
\newblock Cnn lstm network architecture for modeling software reliability.
\newblock In {\em Software Technology: Methods and Tools: 51st International Conference, TOOLS 2019, Innopolis, Russia, October 15--17, 2019, Proceedings 51}, pages 210--217. Springer, 2019.

\bibitem{yangzhen2017software}
Fu~Yangzhen, Zhang Hong, Zeng Chenchen, and Feng Chao.
\newblock A software reliability prediction model: Using improved long short term memory network.
\newblock In {\em 2017 IEEE International Conference on Software Quality, Reliability and Security Companion (QRS-C)}, pages 614--615. IEEE, 2017.

\bibitem{wu2020hybrid}
Maochuan Wu, Junyu Lin, Shouchuang Shi, Long Ren, and Zhiwen Wang.
\newblock Hybrid optimization-based gru neural network for software reliability prediction.
\newblock In {\em International conference of pioneering computer scientists, engineers and educators}, pages 369--383. Springer, 2020.

\bibitem{yang2024software}
Minghao Yang, Shunkun Yang, and Chong Bian.
\newblock Software reliability prediction by adaptive gated recurrent unit-based encoder-decoder model with ensemble empirical mode decomposition.
\newblock {\em Software Testing, Verification and Reliability}, 34(8):e1895, 2024.

\bibitem{kumar2012empirical}
Pradeep Kumar and Yogesh Singh.
\newblock An empirical study of software reliability prediction using machine learning techniques.
\newblock {\em International Journal of System Assurance Engineering and Management}, 3:194--208, 2012.

\bibitem{rani2018neural}
Pooja Rani and GS~Mahapatra.
\newblock Neural network for software reliability analysis of dynamically weighted nhpp growth models with imperfect debugging.
\newblock {\em Software Testing, Verification and Reliability}, 28(5):e1663, 2018.

\bibitem{xie1997practical}
Min Xie, Guan~Y Hong, and Claes Wohlin.
\newblock A practical method for the estimation of software reliability growth in the early stage of testing.
\newblock In {\em PROCEEDINGS The Eighth International Symposium On Software Reliability Engineering}, pages 116--123. IEEE, 1997.

\bibitem{azturk2021ensemble}
Elifnur Azturk and Ayse Tosun.
\newblock Ensemble and cross-project software reliability growth models for safety-critical systems.
\newblock In {\em 2021 6th International Conference on Computer Science and Engineering (UBMK)}, pages 785--790. IEEE, 2021.

\bibitem{zhang2024bootstrap}
Wei Zhang and Jianhui Jiang.
\newblock Bootstrap-based resampling methods for software reliability measurement under small sample condition.
\newblock {\em Journal of Circuits, Systems and Computers}, 33(09):2450161, 2024.

\bibitem{san2021deep}
Kyawt~Kyawt San, Hironori Washizaki, Yoshiaki Fukazawa, Kiyoshi Honda, Masahiro Taga, and Akira Matsuzaki.
\newblock Deep cross-project software reliability growth model using project similarity-based clustering.
\newblock {\em Mathematics}, 9(22):2945, 2021.

\bibitem{nagaraju2022adaptive}
Vidhyashree Nagaraju, Shadow Pritchard, and Lance Fiondella.
\newblock Adaptive incremental learning for software reliability growth models.
\newblock In {\em International Conference on Human-Computer Interaction}, pages 352--366. Springer, 2022.

\bibitem{wood1996predicting}
Alan Wood.
\newblock Predicting software reliability.
\newblock {\em Computer}, 29(11):69--77, 1996.

\bibitem{karunanithi1991prediction}
Nachimuthu Karunanithi, Yashwant~K Malaiya, and L~Darrell Whitley.
\newblock Prediction of software reliability using neural networks.
\newblock In {\em ISSRE}, pages 124--130, 1991.

\bibitem{musa1979software}
John~D Musa.
\newblock Software reliability data.
\newblock {\em technical report, data and analysis center for software, Rome Air Development Center, Griffis AFB}, 1979.

\bibitem{lyu1996handbook}
Michael~R Lyu et~al.
\newblock {\em Handbook of software reliability engineering}, volume 222.
\newblock IEEE computer society press Los Alamitos, 1996.

\bibitem{mivcko2022applicability}
Radoslav Mi{\v{c}}ko, Stanislav Chren, and Bruno Rossi.
\newblock Applicability of software reliability growth models to open source software.
\newblock In {\em 2022 48th Euromicro Conference on Software Engineering and Advanced Applications (SEAA)}, pages 255--262. IEEE, 2022.

\bibitem{kim2024enhancing}
Taehyoun Kim, Duksan Ryu, and Jongmoon Baik.
\newblock Enhancing software reliability growth modeling: A comprehensive analysis of historical datasets and optimal model selections.
\newblock In {\em 2024 IEEE 24th International Conference on Software Quality, Reliability and Security (QRS)}, pages 147--158. IEEE, 2024.

\bibitem{xie1999software}
Min Xie, Guan~Y Hong, and Claes Wohlin.
\newblock Software reliability prediction incorporating information from a similar project.
\newblock {\em Journal of Systems and Software}, 49(1):43--48, 1999.

\bibitem{rana2013evaluating}
Rakesh Rana, Miroslaw Staron, Christian Berger, J{\"o}rgen Hansson, Martin Nilsson, and Fredrik T{\"o}rner.
\newblock Evaluating long-term predictive power of standard reliability growth models on automotive systems.
\newblock In {\em 2013 IEEE 24th International Symposium on Software Reliability Engineering (ISSRE)}, pages 228--237. IEEE, 2013.

\bibitem{honda2016case}
Kiyoshi Honda, Nobuhiro Nakamura, Hironori Washizaki, and Yoshiaki Fukazawa.
\newblock Case study: Project management using cross project software reliability growth model.
\newblock In {\em 2016 IEEE International Conference on Software Quality, Reliability and Security Companion (QRS-C)}, pages 39--46. IEEE, 2016.

\bibitem{hu2006early}
QP~Hu, Yuan-Shun Dai, Min Xie, and SH~Ng.
\newblock Early software reliability prediction with extended ann model.
\newblock In {\em 30th Annual International Computer Software and Applications Conference (COMPSAC'06)}, volume~2, pages 234--239. IEEE, 2006.

\bibitem{san2019dc}
Kyawt~Kyawt San, Hironori Washizaki, Yoshiaki Fukazawa, Kiyoshi Honda, Masahiro Taga, and Akira Matsuzaki.
\newblock Dc-srgm: Deep cross-project software reliability growth model.
\newblock In {\em 2019 IEEE International Symposium on Software Reliability Engineering Workshops (ISSREW)}, pages 61--66. IEEE, 2019.

\bibitem{costa2007exploring}
Eduardo~Oliveira Costa, Gustavo~Alexandre de~Souza, Aurora Trinidad~Ramirez Pozo, and Silvia~Regina Vergilio.
\newblock Exploring genetic programming and boosting techniques to model software reliability.
\newblock {\em IEEE Transactions on Reliability}, 56(3):422--434, 2007.

\bibitem{costa2010genetic}
Eduardo~Oliveira Costa, Aurora Trinidad~Ramirez Pozo, and Silvia~Regina Vergilio.
\newblock A genetic programming approach for software reliability modeling.
\newblock {\em IEEE transactions on reliability}, 59(1):222--230, 2010.

\bibitem{nagaraju2017performance}
Vidhyashree Nagaraju, Lance Fiondella, Panlop Zeephongsekul, Chathuri~L Jayasinghe, and Thierry Wandji.
\newblock Performance optimized expectation conditional maximization algorithms for nonhomogeneous poisson process software reliability models.
\newblock {\em IEEE Transactions on Reliability}, 66(3):722--734, 2017.

\bibitem{el2020practical}
Khaled El~Emam, Lucy Mosquera, and Richard Hoptroff.
\newblock {\em Practical synthetic data generation: balancing privacy and the broad availability of data}.
\newblock O'Reilly Media, 2020.

\bibitem{egri2017cross}
Attila Egri, Ill{\'e}s Horv{\'a}th, Ferenc Kov{\'a}cs, Roland Molontay, and Kriszti{\'a}n Varga.
\newblock Cross-correlation based clustering and dimension reduction of multivariate time series.
\newblock In {\em 2017 IEEE 21st International Conference on Intelligent Engineering Systems (INES)}, pages 000241--000246. IEEE, 2017.

\bibitem{kim2021reversible}
Taesung Kim, Jinhee Kim, Yunwon Tae, Cheonbok Park, Jang-Ho Choi, and Jaegul Choo.
\newblock Reversible instance normalization for accurate time-series forecasting against distribution shift.
\newblock In {\em International conference on learning representations}, 2021.

\bibitem{huang2011estimation}
Chin-Yu Huang and Michael~R Lyu.
\newblock Estimation and analysis of some generalized multiple change-point software reliability models.
\newblock {\em IEEE Transactions on reliability}, 60(2):498--514, 2011.

\bibitem{wu2021study}
Cheng-Yang Wu and Chin-Yu Huang.
\newblock A study of incorporation of deep learning into software reliability modeling and assessment.
\newblock {\em IEEE Transactions on Reliability}, 70(4):1621--1640, 2021.

\bibitem{satoh2000discrete}
Daisuke Satoh.
\newblock A discrete gompertz equation and a software reliability growth model.
\newblock {\em IEICE TRANSACTIONS on Information and Systems}, 83(7):1508--1513, 2000.

\bibitem{jeske2005adjusting}
Daniel~R Jeske, Xuemei Zhang, and Loan Pham.
\newblock Adjusting software failure rates that are estimated from test data.
\newblock {\em IEEE Transactions on Reliability}, 54(1):107--114, 2005.

\bibitem{jeske2005some}
Daniel~R Jeske and Xuemei Zhang.
\newblock Some successful approaches to software reliability modeling in industry.
\newblock {\em Journal of Systems and Software}, 74(1):85--99, 2005.

\bibitem{kenny1993estimating}
Garrison~Q Kenny.
\newblock Estimating defects in commercial software during operational use.
\newblock {\em IEEE Transactions on reliability}, 42(1):107--115, 1993.

\bibitem{kaufman1999using}
Lori~M Kaufman, Joanne~Bechta Dugan, and Barry~W Johnson.
\newblock Using statistics of the extremes for software reliability analysis.
\newblock {\em IEEE Transactions on Reliability}, 48(3):292--299, 1999.

\bibitem{derriennic1995use}
H~Derriennic and G~Le~Gall.
\newblock Use of failure-intensity models in the software-validation phase for telecommunications.
\newblock {\em IEEE transactions on reliability}, 44(4):658--665, 1995.

\bibitem{musa1987software}
JD~Musa.
\newblock Software reliability: Measurement, prediction, application, 1987.

\bibitem{yang1996infinite}
Kune-Zang Yang.
\newblock {\em An infinite server queueing model for software readiness assessment and related performance measures}.
\newblock Syracuse University, 1996.

\bibitem{kaaniche1992discrete}
Mohamed Ka{\^a}niche and Karama Kanoun.
\newblock The discrete time hyperexponential model for software reliability growth evaluation.
\newblock In {\em Third International Symposium on Software Reliability Engineering (ISSRE-1992)}, pages 64--75, 1992.

\bibitem{ohba1982software}
Mitsuru Ohba.
\newblock Software quality= test accuracy$\times$ test coverage.
\newblock In {\em Proceedings of the 6th international conference on Software engineering}, pages 287--293, 1982.

\bibitem{ohba1984software}
Mitsuru Ohba.
\newblock Software reliability analysis models.
\newblock {\em IBM Journal of research and Development}, 28(4):428--443, 1984.

\bibitem{xie2007study}
Min Xie, QP~Hu, YP~Wu, and Szu~Hui Ng.
\newblock A study of the modeling and analysis of software fault-detection and fault-correction processes.
\newblock {\em Quality and Reliability Engineering International}, 23(4):459--470, 2007.

\bibitem{martini1990software}
MRd~Bastos Martini, Karama Kanoun, and J~Moreira de~Souza.
\newblock Software-reliability evaluation of the tropico-r switching system.
\newblock {\em IEEE Transactions on Reliability}, 39(3):369--379, 1990.

\bibitem{misra1983software}
Pratap~N. Misra.
\newblock Software reliability analysis.
\newblock {\em IBM Systems Journal}, 22(3):262--270, 1983.

\bibitem{garg2021decision}
Rakesh Garg, Supriya Raheja, and Ramesh~Kumar Garg.
\newblock Decision support system for optimal selection of software reliability growth models using a hybrid approach.
\newblock {\em IEEE Transactions on Reliability}, 71(1):149--161, 2021.

\bibitem{mullen1998lognormal}
Robert~E Mullen.
\newblock The lognormal distribution of software failure rates: application to software reliability growth modeling.
\newblock In {\em Proceedings Ninth International Symposium on Software Reliability Engineering (Cat. No. 98TB100257)}, pages 134--142. IEEE, 1998.

\bibitem{brooks1980analysis}
WD~Brooks and RW~Motley.
\newblock Analysis of discrete software reliability models.
\newblock {\em (No Title)}, 1980.

\bibitem{li2011reliability}
Xiang Li, Yan~Fu Li, Min Xie, and Szu~Hui Ng.
\newblock Reliability analysis and optimal version-updating for open source software.
\newblock {\em Information and Software Technology}, 53(9):929--936, 2011.

\bibitem{zhang2002calibrating}
Xuemei Zhang, Daniel~R Jeske, and Hoang Pham.
\newblock Calibrating software reliability models when the test environment does not match the user environment.
\newblock {\em Applied Stochastic Models in Business and Industry}, 18(1):87--99, 2002.

\bibitem{zhang2006software}
Xuemei Zhang and Hoang Pham.
\newblock Software field failure rate prediction before software deployment.
\newblock {\em Journal of Systems and Software}, 79(3):291--300, 2006.

\bibitem{tohma1989structural}
Yoshihiro Tohma, Kenshin Tokunaga, Shinji Nagase, and Yukihisa Murata.
\newblock Structural approach to the estimation of the number of residual software faults based on the hyper-geometric distribution.
\newblock {\em IEEE transactions on software engineering}, 15(3):345--355, 1989.

\bibitem{tohma1989hyper}
Yoshihiro Tohma, Raymond Jacoby, Yukihisa Murata, and Moriki Yamamoto.
\newblock Hyper-geometric distribution model to estimate the number of residual software faults.
\newblock In {\em [1989] Proceedings of the Thirteenth Annual International Computer Software \& Applications Conference}, pages 610--617. IEEE, 1989.

\bibitem{tohma1991parameter}
Yoshihiro Tohma, Hisashi Yamano, Morio Ohba, and Raymond Jacoby.
\newblock Parameter estimation of the hyper-geometric distribution model for real test/debug data.
\newblock In {\em Proceedings. 1991 International Symposium on Software Reliability Engineering}, pages 28--29. IEEE Computer Society, 1991.

\bibitem{yamada1985software}
Shigeru Yamada and Shunji Osaki.
\newblock Software reliability growth modeling: Models and applications.
\newblock {\em IEEE Transactions on software engineering}, (12):1431--1437, 1985.

\bibitem{wilcoxon1992individual}
Frank Wilcoxon.
\newblock Individual comparisons by ranking methods.
\newblock In {\em Breakthroughs in statistics: Methodology and distribution}, pages 196--202. Springer, 1992.

\bibitem{friedman1937use}
Milton Friedman.
\newblock The use of ranks to avoid the assumption of normality implicit in the analysis of variance.
\newblock {\em Journal of the american statistical association}, 32(200):675--701, 1937.

\end{thebibliography}

\appendix
\section*{Appendix}
\section{Selection of Best Traditional SRGM}\label{appendix:best_srgm}

To establish the best traditional SRGM baseline, we evaluate six widely used models on 60 real-world defect datasets. These include four traditional SRGMs and two general-purpose statistical growth models. Although the L and G models were not originally designed for software reliability modeling, they have been widely adopted in prior SRGM studies due to their flexible growth curve characteristics and empirical effectiveness. Their mathematical definitions and types are summarized in Table~\ref{tab:baseline_models_appendix}. 

\begin{table}[tbh]
\centering
\caption{Mathematical formulations of models considered for best traditional SRGM selection.}
\label{tab:baseline_models_appendix}
\scriptsize
\begin{tabular}{lccc}
\toprule
\textbf{Model} & \textbf{Type} & \textbf{Equation $m(t)$} & \textbf{Reference} \\
\midrule
Goel-Okumoto (GO) & Concave &
$m(t) = a \left(1 - e^{-bt}\right), \quad 0 < a,\; 0 < b < 1$ & \cite{goel1979time} \\
Yamada Delayed S-Shaped (YDSS) & S-shaped &
$m(t) = a \left(1 - (1 + bt)e^{-bt}\right), \quad 0 < a,\; 0 < b < 1$ & \cite{yamada2009s} \\
Inflection S-shaped (ISS) & S-shaped &
$m(t) = a \left( \frac{1 - e^{-bt}}{1 + \frac{1 - r}{r}e^{-bt}} \right), \quad 0 < a,\; 0 < b < 1,\; 0 < r < 1$ & \cite{ohba1984inflection} \\
Generalized Goel (GG) & Concave/S-shaped &
$m(t) = a \left(1 - e^{-bt}\right)^c, \quad 0 < a,\; 0 < b < 1,\; 0 < c$ & \cite{goel1982software} \\
Logistic (L) & Concave/S-shaped & $m(t) = \frac{a}{1 + ce^{-bt}}, \quad 0 < a,\; 0 < b,\; 0 < c$ & \cite{yamada1985software} \\
Gompertz (G) & S-shaped & $m(t) = a b^{c^{t}}, \quad 0 < a,\; 0 < b < 1,\; 0 < c < 1$ & \cite{yamada1985software} \\
\bottomrule
\end{tabular}
\end{table}

For each dataset, the first 50\% of the data is used to fit all six SRGMs. Model parameters are estimated using the \texttt{curve\_fit} function in Python with the least squares optimization method. Table~\ref{tab:best_srgm_results} reports the MSE values of all six models evaluated on the first 50\% interval for each dataset. If parameter estimation fails for a particular model on a given dataset, the corresponding MSE is recorded as \texttt{Inf}, indicating that the model could not be successfully fitted. The model with the lowest MSE is selected as the best SRGM, and its abbreviation is recorded in the final column. The lowest MSE value in each row is highlighted in bold.

\begin{sidewaystable}[p]
\caption{MSE values for six traditional SRGMs fitted to the first 50\% of each dataset.}%
\label{tab:best_srgm_results}
\scriptsize
\begin{tabular}{lrrrrrrl|lrrrrrrl}%
\toprule
  \textbf{ID} & \textbf{GO} & \textbf{YDSS} & \textbf{ISS} & \textbf{GG} & \textbf{L} & \textbf{G} & \textbf{Best}& \textbf{ ID} & \textbf{GO} & \textbf{YDSS} & \textbf{ISS} & \textbf{GG} & \textbf{L} & \textbf{G} & \textbf{Best}\\
\midrule
DS 1 & 9.9667& 34.0546& 9.9667& 259.7556& 2.3942& \textbf{2.0399}& G& DS 67& 34.4946& 7.8489& 1.8183& 4.5720& \textbf{1.1181}& 3.0147& L\\
DS 2 & Inf& 27.2217& Inf& Inf& 2.1260& \textbf{2.0078}& G& DS 78& 17.3855& 16.2069& 15.8416& 15.6561& 16.4232& \textbf{14.8829}& G\\
DS 4 & Inf& 1.2517& \textbf{0.9567}& Inf& 1.2354& 1.0889& ISS& DS 79& Inf& 27.5583& \textbf{21.5109}& Inf& 27.7991& 23.6397& ISS\\
DS 8 & Inf& 2282.3028& \textbf{301.2440}& Inf& 1078.9813& 611.5731& ISS& DS 80& Inf& 241.5723& Inf& Inf& 227.7184& \textbf{178.8044}& G\\
DS 9 & Inf& 2820.0832& \textbf{321.3838}& Inf& 1203.1006& 687.2489& ISS& DS 81& Inf& 9.0363& 10.3387& 9.1152& 11.4830& \textbf{8.9937}& G\\
DS 10& Inf& 5107.7882& \textbf{581.6245}& Inf& 3188.2560& 1713.1986& ISS& DS 82& 67.6718& \textbf{10.3303}& 21.5462& 12.4340& 48.6648& 18.7392& YDSS    \\
DS 11& Inf& 4426.0344& \textbf{550.6592}& Inf& 2978.9751& 1575.1637& ISS& DS 83& Inf& 3.6349& 1.4830& 0.7064& 2.8106& \textbf{0.5353}& G\\
DS 12& Inf& Inf& Inf& 905.5556& \textbf{80.2587}& Inf& L& DS 85& Inf& 56.2068& 21.0981& 15.9157& 26.5380& \textbf{13.1052}& G\\
DS 13& 126.2139& 350.0426& 126.2139& 854.2143& \textbf{23.2847}& 29.3801& L& DS 89& 95.5868& 246.2935& \textbf{87.3510}& 92.9183& 135.1149& 107.4546& ISS\\
DS 15& 11.7443& 39.2474& 11.7443& Inf& 5.0397& \textbf{4.0944}& G& DS 91& 17.2488& 73.8893& 17.2488& Inf& 15.7237& \textbf{10.8458}& G\\
DS 16& 127471.65& \textbf{63761.64}& 69919.01& 68901.15& 82566.21& 64323.87& YDSS    & DS 92& 94.18& 375.15& 94.18& 3092.87& 85.41& \textbf{66.93}& G\\
DS 17& Inf& 134.3751& \textbf{92.7317}& 121.1463& 332.7338& 138.7442& ISS& DS 94& Inf& 14.7908& 10.4271& \textbf{8.7479}& 45.0561& 17.3004& GG\\
DS 18& 763306.97& 294195.66& 336375.71& 321985.03& 406748.59& \textbf{263277.02}& G& DS 95& 71.9336& 5.9658& 6.6384& \textbf{3.9463}& 19.8735& 4.4159& GG\\
DS 19& 8086.40& 17722.98& 6855.73& \textbf{5588.42}& 25902.80& 12445.22& GG& DS 97& Inf& 411.4606& \textbf{95.1849}& Inf& 267.3901& 174.2560& ISS\\
DS 25& Inf& 12.1601& 6.5741& \textbf{6.3891}& 11.2807& 7.8780& GG& DS 106& 3.6481& 10.5790& 3.6481& \textbf{3.5335}& 4.5326& 3.6095& GG\\
DS 26& Inf& 0.1709& 0.0766& \textbf{0.0743}& 0.1584& 0.1104& GG& DS 107& 6.4493& 15.6487& 6.4493& Inf& 5.1610& \textbf{5.0406}& G\\
DS 27& 15.7059& 4.8579& \textbf{3.1743}& 4.0360& 5.7121& 3.5596& ISS& DS 108& 2.1823& 4.5615& \textbf{1.1567}& 1.4281& 2.1938& 1.1765& ISS\\
DS 28& Inf& 11.4775& 12.8153& \textbf{8.9871}& 28.8992& 13.7058& GG& DS 109& 5.0201& 6.7421& 3.6612& 4.6895& \textbf{1.6392}& 2.8687& L\\
DS 29& 20.5395& 52.7400& 20.5395& \textbf{20.2480}& 27.0696& 20.4379& GG& DS 110& 4.4678& 1.8216& 1.4992& 1.6838& 1.6411& \textbf{1.3691}& G\\
DS 30& 4.1714& 9.0164& 4.1714& \textbf{4.0229}& 6.5919& 5.3265& GG& DS 111& 3.0555& 6.6967& 3.0503& \textbf{3.0116}& 5.7828& 4.0520& GG\\
DS 47& 17.5708& 5.0543& 6.6338& 5.4004& 9.0133& \textbf{4.6354}& G& DS 114& Inf& 0.00031& \textbf{0.00017}& 0.00022& 0.00042& 0.00024& ISS\\
DS 48& Inf& 0.2384& 0.2157& 0.2360& \textbf{0.2079}& 0.2183& L& DS 115& Inf& 0.00005& 0.00005& \textbf{0.00002}& 0.00023& 0.00008& GG\\
DS 50& Inf& Inf& 4.0903& 4.5581& \textbf{4.0249}& 5.6923& L& DS 116& Inf& 0.0009& Inf& Inf& \textbf{0.0003}& Inf& L\\
DS 57& Inf& Inf& 528.04& 59986.48& \textbf{517.73}& 1653.05& L& DS 117& Inf& 0.001285& Inf& Inf& 0.000394& \textbf{0.000389}& G\\
DS 58& 2.8725& 4.6283& 2.8725& \textbf{2.8248}& 4.6957& 3.5604& GG& DS 121& 449.2898& 586.7416& 447.0016& \textbf{427.9786}& 807.4659& 593.8310& GG\\
DS 60& Inf& 1.0668& 0.4491& 0.2830& \textbf{0.2678}& 0.3864& L& DS 122& Inf& 3.8628& 2.1663& 18.1961& \textbf{2.0938}& 2.8221& L\\
DS 61& Inf& 1.7585& Inf& Inf& 0.6790& \textbf{0.6774}& G& DS 123& Inf& 26.1129& Inf& Inf& 17.1513& \textbf{14.6866}& G\\
DS 63& 80.8790& 122.4204& 80.8790& \textbf{80.4837}& 122.7603& 101.0899& GG& DS 124& Inf& 8.1605& Inf& Inf& 8.4137& \textbf{6.2933}& G\\
DS 65& 127.8828& 39.2422& 45.6895& \textbf{38.2662}& 69.1232& 44.0872& GG& DS 125& 105.5704& 96.0972& 98.3856& \textbf{92.7174}& 128.8746& 104.7903& GG\\
DS 66& 173.0061& 60.2740& \textbf{36.9866}& 43.8164& 39.8143& 46.5192& ISS& DS 126& & 279.1850& Inf& Inf& 197.4702& \textbf{110.4127}& G \\
\bottomrule
\end{tabular}
\end{sidewaystable}

An analysis of Table~\ref{tab:best_srgm_results} reveals that the G model was the most frequently selected traditional SRGM, achieving the lowest MSE in 19 out of the 60 datasets. It was followed by the GG model, which was selected as the best-fitting model in 16 datasets. The ISS and L models were chosen in 13 and 10 datasets, respectively. The YDSS model was selected for only two datasets, while the GO model was not selected as the best model for any dataset.

These results suggest that the general-purpose statistical models, namely the G and L models, are highly effective at capturing the defect discovery patterns observed in the benchmark datasets. Although the GO model has historically been widely used in SRGM studies, its absence among the models with the lowest MSE indicates limited flexibility in modeling diverse failure trends. However, it is important to note that a model showing the best fit on the initial 50\% of the data does not necessarily guarantee superior predictive performance on the remaining portion.

\section{Additional Scatter Plot Visualizations}\label{appendix:scatter}

This appendix presents additional scatter plots for RMSE and MAPE corresponding to RQ1, RQ2, and RQ3. All metrics are log-scaled to enhance visual comparability and mitigate the impact of extreme values. For MAE plots, refer to Figures~\ref{fig:rq1_scatter},~\ref{fig:rq2_scatter}, and~\ref{fig:rq3_scatter} in the main text.

\begin{figure}[tbh]
\centering
\includegraphics[width=0.97\textwidth]{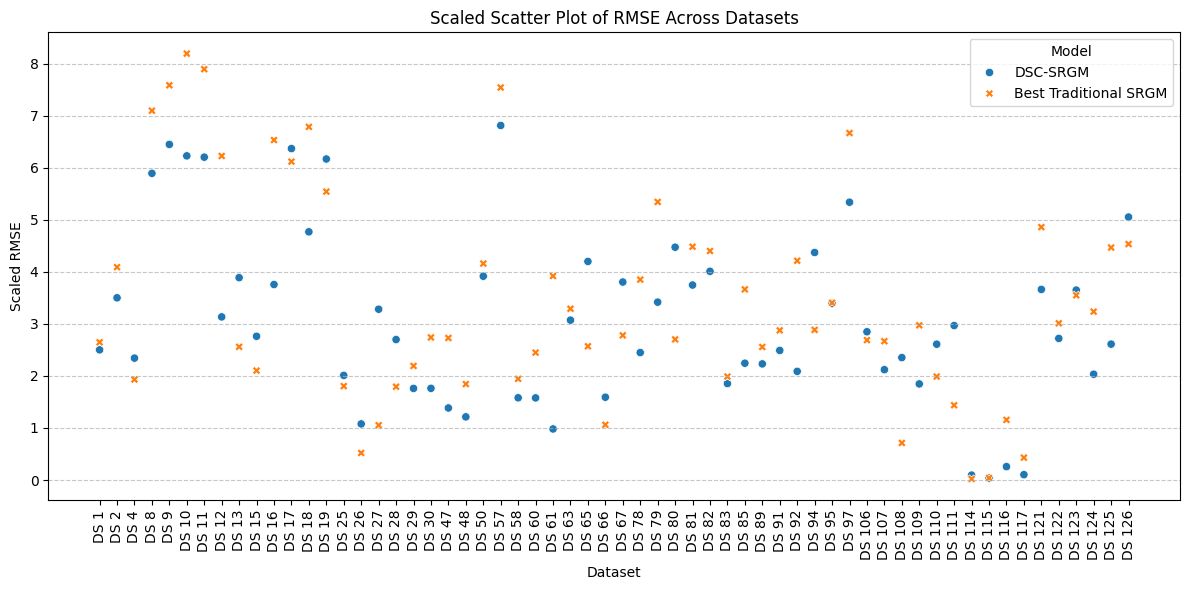}
\caption{RQ1 – Scatter plot of log-scaled RMSE.}
\label{fig:rq1_rmse_scatter}
\end{figure}

\begin{figure}[tbh]
\centering
\includegraphics[width=0.97\textwidth]{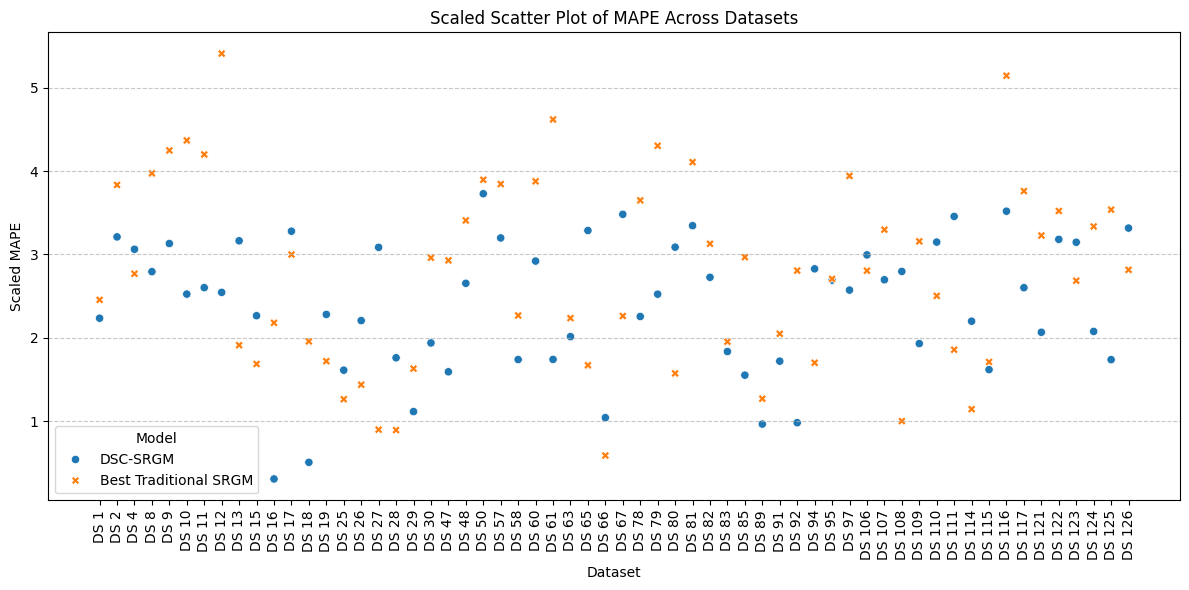}
\caption{RQ1 – Scatter plot of log-scaled MAPE.}
\label{fig:rq1_mape_scatter}
\end{figure}

\begin{figure}[tbh]
\centering
\includegraphics[width=\textwidth]{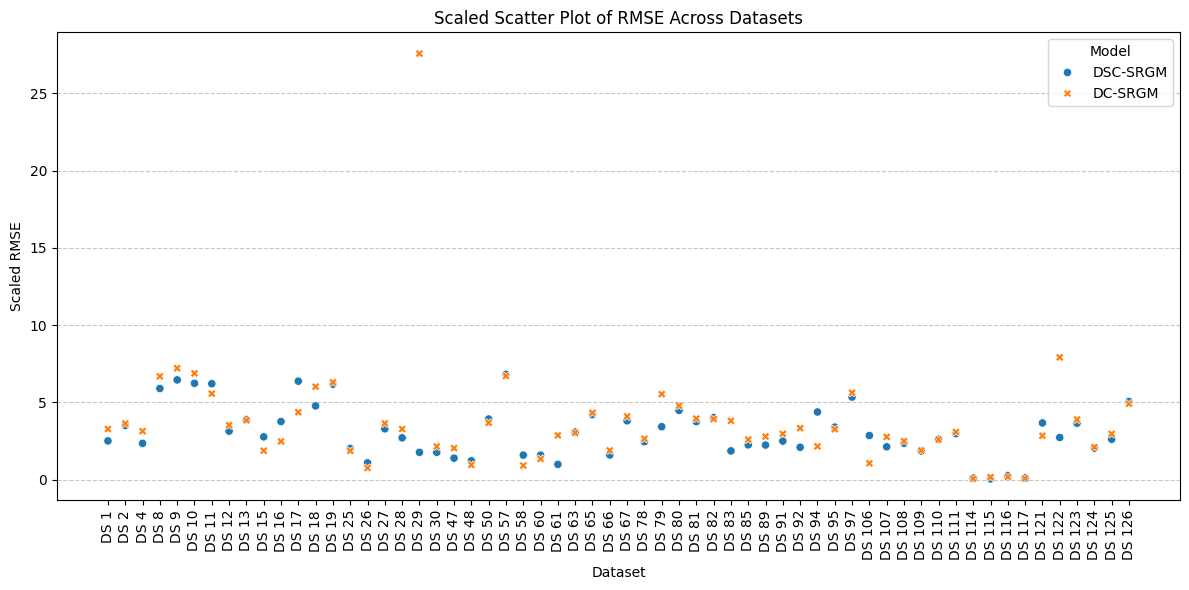}
\caption{RQ2 – Scatter plot of log-scaled RMSE.}
\label{fig:rq2_rmse_scatter}
\end{figure}

\begin{figure}[tbh]
\centering
\includegraphics[width=\textwidth]{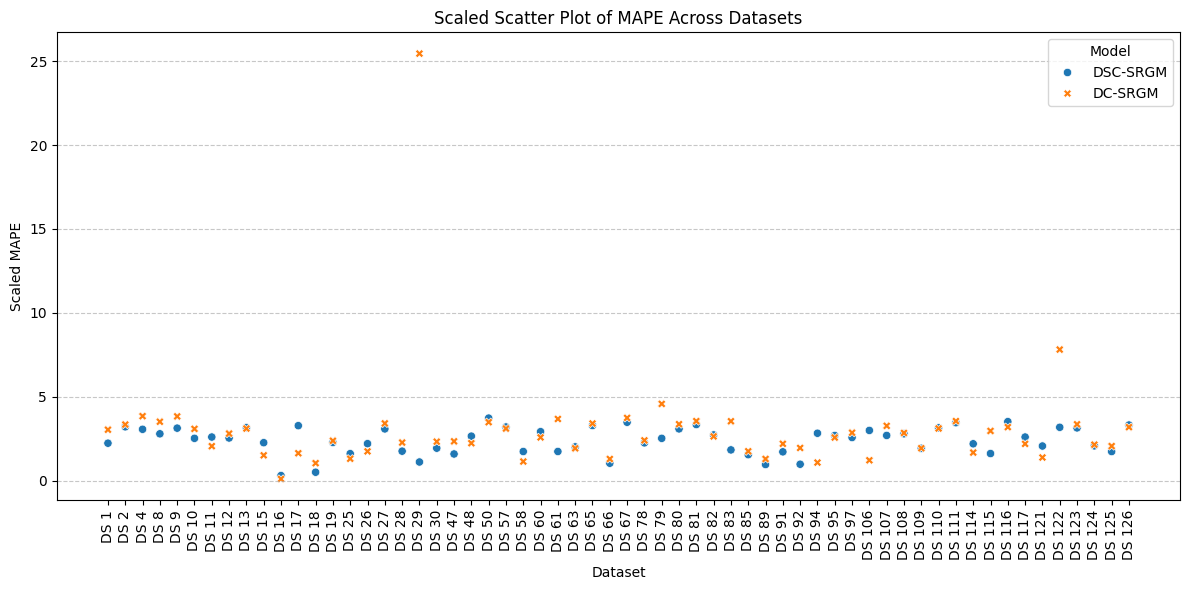}
\caption{RQ2 – Scatter plot of log-scaled MAPE.}
\label{fig:rq2_mape_scatter}
\end{figure}

\begin{figure}[tbh]
\centering
\includegraphics[width=\textwidth]{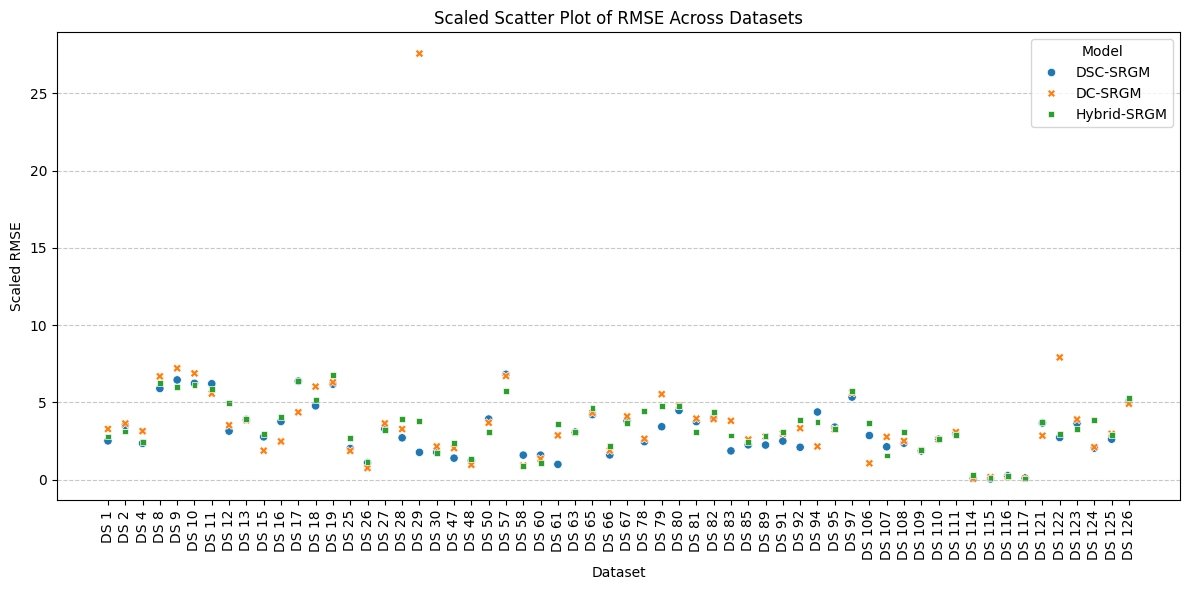}
\caption{RQ3 – Scatter plot of log-scaled RMSE.}
\label{fig:rq3_rmse_scatter}
\end{figure}

\begin{figure}[tbh]
\centering
\includegraphics[width=\textwidth]{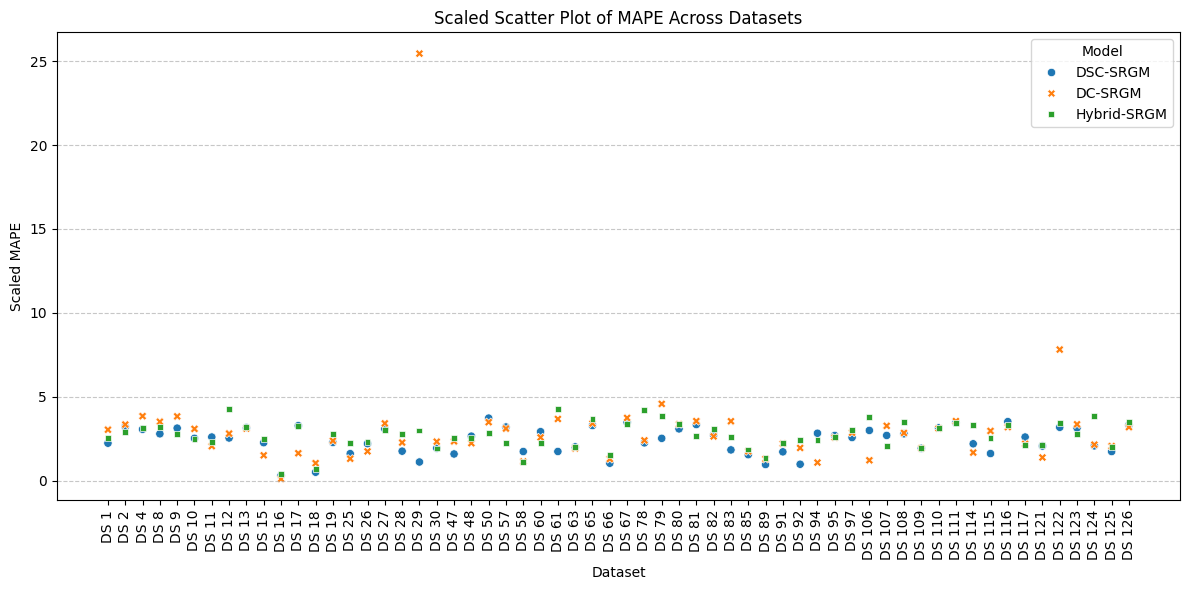}
\caption{RQ3 – Scatter plot of log-scaled MAPE.}
\label{fig:rq3_mape_scatter}
\end{figure}

\end{document}